\documentstyle[prl,aps,multicol,eqsecnum]{revtex}

\begin{document}
\draft

\def\beq{\begin{equation}}
\def\eeq{\end{equation}}
\def\eeql#1{\label{#1} \end{equation}}
\def\bea{\begin{eqnarray}}
\def\eea{\end{eqnarray}}
\def\eeal#1{\label{#1} \end{eqnarray}}
\def\phs{{\vphantom{*}}}
\def\hn{\mskip-0.5\thinmuskip}
\def\hp{\mskip0.5\thinmuskip}
\renewcommand{\vec}{\bbox}
\def\im{\mathop{\rm Im}\nolimits}
\def\re{\mathop{\rm Re}\nolimits}
\def\ds{\displaystyle}
\def\ts{\textstyle}
\def\half{{\ts\frac{1}{2}}}
\def\lbar{{\mathchar '26\mkern -9mu\lambda}}
\def\abs#1{\mathopen|#1\mathclose|}

\title{Analytic treatment of black-hole gravitational waves\\ at the algebraically special frequency}

\author{Alec Maassen van den Brink\cite{elad}}

\address{Department of Physics, The Chinese University of Hong Kong, Hong Kong, China {\em and}\\ Institute of Theoretical Physics, Chinese Academy of Sciences,
Beijing, China}

\date{\today}
\maketitle

\begin{abstract}

We study the Regge--Wheeler and Zerilli equations (RWE and ZE) at the `algebraically special frequency' $\Omega$, where these equations admit an exact solution (elaborated here), generating the SUSY relationship between them. The physical significance of the SUSY generator and of the solutions at $\Omega$ in general is elucidated as follows. The RWE has no (quasinormal or total-transmission) modes at all; however, $\Omega$ is nonetheless `special' in that (a) for the outgoing wave into the horizon one has a `miraculous' cancellation of a divergence expected due to the exponential potential tail, and (b) the branch-cut discontinuity at $\omega=\Omega$ vanishes in the outgoing wave to infinity. Moreover, (a) and (b) are related. For the ZE, its only mode is the---inverse---SUSY generator, which is at the same time a quasinormal mode {\em and\/} a total-transmission mode propagating to infinity. The subtlety of these findings (of general relevance for future study of the equations on or near the negative imaginary $\omega$-axis) may help explain why the situation has sometimes been controversial. For finite black-hole rotation, the algebraically special modes are shown to be totally transmitting, and the implied singular nature of the Schwarzschild limit is clarified. The analysis draws on a recent detailed investigation of SUSY in open systems [math-ph/9909030].

\end{abstract}

\pacs{PACS numbers: 04.30.-w
, 04.20.Jb
, 04.70.Bw
, 11.30.Pb
}
 
\begin{multicols}{2}

\section{Introduction}
\label{intro}

The evolution of small perturbations of the Schwarz\-schild metric of a spherically symmetric black hole, such as generated by infalling matter\cite{davis} or in the aftermath of a stellar collapse\cite{gaiser}, is well known\cite{grav} to be described in each angular-momentum sector $\ell\ge2$ by a Klein--Gordon wave equation
\beq
  [\partial_x^2+\omega^2-V(x)]\hp\phi(x,\omega)=0\;.
\eeql{KGE}
Here, $\phi$ is a scalar field (a combination of the metric-function perturbations), while the `tortoise coordinate' is
\beq
  x=r+\ln(r-1)\;,
\eeql{x-s}
where $r$ is the circumferential radius (so that $1<r<\infty$ maps to $-\infty<x<\infty$), and $c=G=2M=1$ ($M$ is the black-hole mass). However, the axial sector is described by the Regge--Wheeler equation (RWE), which has the potential
\beq
  V(x)=(r-1)\frac{2(n{+}1)r-3}{r^4}\;,
\eeql{VRW}
while in the polar sector, the Zerilli equation (ZE) has
\beq
  \tilde{V}(x)=(r-1)\frac{8n^2(n{+}1)r^3+12n^2r^2+18nr+9}{r^4(2nr+3)^2}\;,
\eeql{VZ}
where we defined $n=\half(\ell{-}1)(\ell{+}2)$.

Clearly, one is now interested in the {\em modes\/} of these equations, i.e., eigenfunctions under either outgoing- ($\phi(x{\rightarrow}{\pm}\infty,\omega)\sim e^{i\omega|x|}$; however, see below for precise definitions) or incoming-wave ($\phi(x,\omega)\sim e^{-i\omega|x|}$) boundary conditions\cite{out-in}. Foremost among these are the {\em quasinormal modes\/} (QNMs, with outgoing boundary conditions), which constitute the excitation spectrum of the black hole. Besides being a matter of principle in deciding the stability of the Schwarzschild solution, the issue takes on exciting significance with the prospect of their experimental detection in the coming decade using instruments such as LIGO and VIRGO\cite{ligo}. QNMs also play a role in fully nonlinear numerical simulations\cite{anninos}, where they are observed to dominate the radiated signal at intermediate times after a violent event such as a stellar collapse or a black-hole collision (cf.\ the beginning of this section), so that numerical and perturbative studies are complementary. Besides QNMs, we will also consider {\em total-transmission modes\/}, incoming from the left (right) but outgoing to the other side, and denoted as TTM$_{\rm L}$s (TTM$_{\rm R}$s). The fourth type of mode, a normal mode (NM) or bound state with $\phi(x,\omega)\sim e^{-|\omega x|}$, is ruled out since both $V$ and $\tilde{V}$ are purely repulsive.

It has been noticed\cite{chanddet} that the RWE and ZE have the same QNM spectrum, which {\em a priori\/} is by no means obvious given the very different forms for the potentials. This can be understood because solutions of the two equations are related by `intertwining'\cite{price} or supersymmetry (SUSY)\cite{susy} as $\tilde{\phi}=[d_x+W(x)]\phi$, with the superpotential
\beq
  W(x)=\frac{N}{2}+\frac{3(r-1)}{r^2(2nr+3)}\;,
\eeql{defW}
where
\beq
  N\equiv\frac{4n(n+1)}{3}=8\left(\!\!\begin{array}{c} \ell+2 \\ 4
  \end{array}\!\!\right)\;.
\eeql{defN}
Of particular interest are now the {\em SUSY generator}
\beq
  \xi_1(x)=\exp\biggl\{-\int^x\!\!dy\,W(y)\biggr\}
\eeql{SUSYgen}
(cf.\ Section~\ref{solveRWE} for the notation) and its counterpart $\tilde{\xi}_1^\phs\equiv\xi_1^{-1}$\cite{tilde}, which generates the inverse transform. By the general theory of SUSY, these are solutions of the RWE and the ZE respectively, at the same eigenvalue, which in this case is yielded by (\ref{KGE}) as $\omega^2=\Omega^2$ for the `algebraically special frequency' $\Omega\equiv-iN\hn/2$.

Since the early study of `algebraically special' black-hole perturbations, these have been associated with TTMs: Ref.\cite{couch}, dealing with the Schwarzschild case, states (below its last, unnumbered display) that `The radiation field (incoming) \ldots\ vanishes.' The analysis of\cite{couch} was generalized to Kerr holes in\cite{wald}, still on the level of general relativity, and subsequently Ref.\cite{chandra} gave the corresponding exact solutions of the RWE, ZE, and Teukolsky equations. Given the fact that\cite{chandra} consistently calls these solutions TTMs, it is remarkable that their solving the appropriate wave equations is checked explicitly, but {\em not\/} their obeying the corresponding boundary conditions.

The situation was confounded when Ref.\cite{leaver1} reported a QNM $\Omega'$ of the RWE numerically very close to, but not at, $\Omega$. For $\ell=3$, however, $\Omega'$ is given with a nonzero real part, while the QNM $-\Omega^{\prime*}$ which would then be dictated by symmetry is not found; this seems to imply that the last 3--4 digits (the order of magnitude of the reported $\re\Omega'$) given by\cite{leaver1} are all insignificant, so that no conclusions can be drawn about possible coincidence of $\Omega$ and $\Omega'$ (if the latter does exist at all\cite{NM}). In contrast, Ref.\cite{and} reports TTM$_{\rm R}$s of the RWE very close to $\Omega$ for $2\le\ell\le6$, and suggests that these correspond to the special mode; again, if all the given decimals are significant, strictly speaking this suggestion is not warranted. Finally, Ref.\cite{onozawa} finds that for $\ell=2$ and for each `magnetic quantum number' $-2\le m\le2$ the Kerr hole has a QNM (the ninth) which tends to $\Omega=-4i$ in the Schwarzschild limit $a\rightarrow0$ ($a$ is the black-hole rotation). However, Ref.\cite{onozawa} cautions that these QNMs should disappear for $a=0$, since they cannot coexist at $\Omega$ with the TTM furnished by the special mode. For each $m$, this disappearance is thought to occur by cancellation of the two modes (mirror images only for $m=0$) on each side of the negative imaginary axis (NIA) in the $\omega$-plane. Certainly this is conceivable, since an analogous cancellation happens for the P\"oschl--Teller potential $V(x)={\cal V}\cosh^{-2}(x)$ (sometimes used as a model for gravitational potentials) for $\sqrt{\frac{1}{4}-{\cal V}}=\frac{1}{2},\frac{3}{2},\frac{5}{2},\ldots$ (e.g.,\cite{susy}).

While the mode situation at $\Omega$ for the RWE alone thus is unclear already, SUSY leaves its relationship to the ZE nontrivial at $\omega^2=\Omega^2$\cite{susy}, $\xi_1$ spanning the kernel of the SUSY operator $d_x+W$. It should be noted that the existing literature rarely emphasizes this potential difference in spectrum between the two equations: for instance, the captions to Figs.\ 1 and~2 and Table~1 of\cite{leaver1} merely say `Schwarzschild QNMs,' although the accompanying text makes it clear that in fact the RWE has been studied. In view of this state of affairs, it seems imperative to start from first principles, using comparatively rigorous methods, and study each of the six questions: at $\Omega$, do the RWE and the ZE have a QNM and/or TTM$_{\rm R}$ and/or TTM$_{\rm L}$? If this investigation can clarify the pitfalls awaiting a numerical analysis, it will be of value beyond this immediate scope.

We will settle the issue as follows: {\em at $\Omega$, the RWE has no modes at all, while the ZE has one mode $\tilde{\xi}_1$ which is simultaneously a QNM and a TTM$_{\rm L}$}. To this end, in Section~\ref{prelim} we first describe some intricacies in the general analysis of (\ref{KGE}) if $V(x)$ is not finitely supported (as would be natural to assume, e.g., in models of cavity QED\cite{RMP}), and set the stage by giving the exact solution of the RWE at $\Omega$. The former (Section~\ref{tails}) is necessarily treated in detail, since the subsequent analysis will turn out to involve precisely the most subtle aspects of the theory. However, readers familiar with the issue, or only interested in results on black holes, can initially skip over this part. In Section~\ref{horizon}, the RWE is studied on the left (near $r=1$). It is found that the SUSY generator (\ref{defW}), (\ref{SUSYgen}) is {\em not\/} outgoing into the horizon, as one might have been tempted to assume; the consequences for SUSY are discussed. A logical question---which will be answered in the negative---is whether any of the physically relevant solutions encountered in the analysis would be distinguished by its behavior near the black-hole singularity $r=0$. Since this involves methods similar to the ones at $r=1$, and since some of the results are relevant to Section~\ref{infty} (there even exists a logical connection back to Section~\ref{horizon}, but that can be established only in Section~\ref{sec:leaver}), the analysis near $r=0$ is presented in Section~\ref{sing}. In Section~\ref{infty}, the RWE is studied on the right (near $r=\infty$); combination of the results of Sections~\ref{horizon}--\ref{infty} will then prove the above claims. In Section~\ref{sec:leaver}, our results are compared and combined with the Leaver-series solution\cite{leaver2} for the outgoing wave to infinity.

Section~\ref{kerr} discusses the special modes of Kerr holes; for $a>0$, these are all found to be TTMs, with both a TTM$_{\rm L}$ and a TTM$_{\rm R}$ (depending on the sign of the spin) present in the lower half $\omega$-plane for each $(\ell,m,a)$. Hence, for the TTM$_{\rm R}$ this property is not conserved in the Schwarzschild limit (since neither the RWE nor the ZE has one, as has been mentioned above), and the reason for this is elucidated. As a consequence of the analysis, it is found that the Schwarzschild special frequency is a limit point not only for the aforementioned TTMs, but for a multiplet of QNMs as well. Finally, Section~\ref{discuss} contains concluding remarks, discussing both the literature cited above and some remaining questions.

\section{Preliminaries}
\label{prelim}

\subsection{Outgoing waves for long-range potentials}
\label{tails}

While it is well known that the tails of the potentials (\ref{VRW}), (\ref{VZ}) make the determination of their modes very hard numerically unless at least $|\im\omega|\lesssim|\re\omega|$, not all authors have sufficiently emphasized several difficulties of principle which occur on the NIA (so-called zero-modes), as already illustrated by the surprising coincidence QNM=TTM$_{\rm L}$ at the end of Section~\ref{intro}. Define the outgoing waves to the left ($f$) and right ($g$) as solutions of (\ref{KGE}) obeying
\beq
  f(x{\rightarrow}{-}\infty,\omega)\sim1\cdot e^{-i\omega x}\;,\quad
  g(x{\rightarrow}\infty,\omega)\sim1\cdot e^{i\omega x}
\eeql{deffg}
for $\im\omega>0$ and further by analytic continuation; these are the functions figuring in the retarded Green's function $\tilde{G}^{\rm R}(x,y;\omega)$. The continuation conserves the asymptotic forms (\ref{deffg}) for those $\omega$ for which the functions are finite, as long as $|\pi/2-\arg(\omega)|<3\pi/2$. A solution is said to be incoming at $\omega$ iff it is outgoing at $-\omega$. The situation, as recently examined by us in\cite{susy} in the context of SUSY, now is as follows.

{\em First}, it could happen that some modes are of higher order, that is, under a generic perturbation they would be split up into several modes rather than merely shifted\cite{jordan,openq}. However, since it will not occur in the present work this possibility will not be pursued here.

{\em Second}, in general the outgoing waves will have a branch point at $\omega=0$ if the tail is stronger than exponential. Eqs.~(\ref{x-s})--(\ref{VZ}) show that this happens for $g$ but not for~$f$; now the NIA is the only choice for the branch cut which respects the symmetry $g(-\omega^*)=g(\omega)^*$. On the cut, one defines $g_{\rm l(r)}(\omega)=\lim_{\epsilon\downarrow0}g(\omega\mp\epsilon)$, and $\delta g(\omega)\equiv g_{\rm r}(\omega)-g_{\rm l}(\omega)$. Since (\ref{deffg}) shows that the increasing parts of $g_{\rm l/r}$ (if present at all, see below) cancel in $\delta g$, and since for each non-real $\omega$ the KGE (\ref{KGE}) has a unique decreasing solution, one must have proportionality
\beq
  \delta g(x,\omega)=\alpha(\omega)\hp g(x,-\omega)\;,
\eeql{def-alpha}
with $\alpha$ purely imaginary\cite{alphaOK}. For each $\omega'$ one further defines a {\em discontinuity index} $\mu$ by $\delta g(\omega)\sim(\omega-\omega')^\mu$. Generically one expects $\mu=0$ while no cut at all would amount to $\mu=\infty$, but at isolated frequencies, corresponding to the zeros of $\alpha$, positive integer values for $\mu$ are possible as well. For further developments, see Appendix~\ref{alpha-app}.

{\em Third}, if the potentials are not oscillating then frequencies $\omega'=-i|\omega'|$ are the only candidates for {\em anomalous points}, where
\beq
  f(\omega)=\frac{{\sf f}(\omega)}{(\omega-\omega')^\nu}\;,
\eeql{f-phi}
with ${\sf f}(\omega)$ finite and nonzero near $\omega=\omega'$ and with the {\em anomalous-point index} $\nu\ge1$ (of course the same can happen for $g$, but this will not occur in the present paper). While these divergences in the outgoing wave are reasonably well known, up to now it apparently has not been realized that they are a mere artifact of the normalization of $f$, so that for instance in $\tilde{G}^{\rm R}(\omega')$ they would cancel against the corresponding divergence in the normalizing Wronskian\cite{RMP}. Rather, the true outgoing solution to the left at the anomalous point is ${\sf f}(\omega')$. Since (\ref{deffg}) and (\ref{f-phi}) show that ${\sf f}(\omega)\sim(\omega-\omega')^\nu e^{-i\omega x}$, it follows that the increasing part of ${\sf f}(\omega')$ vanishes. Hence, ${\sf f}(\omega')$ is actually decreasing for $x\rightarrow-\infty$ and thus coincides with the incoming wave: ${\sf f}(\omega')\propto f(-\omega')$. Because of this counterintuitive behavior, ${\sf f}(\omega')$ is called {\em anomalous outgoing}. It should be noted that $\lim_{\omega\rightarrow\omega'}\lim_{x\rightarrow-\infty}|{\sf f}(x,\omega)|=\infty$ but $\lim_{x\rightarrow-\infty}\lim_{\omega\rightarrow\omega'}|{\sf f}(x,\omega)|=0$, severely hampering any brute-force numerical analysis. However, ${\sf f}(x,\omega)$ is completely regular near $\omega'$ in any finite region of $x$.

Physically, the above merely means that at the anomalous point $\omega'$, the small (typically exponential) tail scatters so strongly that the outgoing wave is completely different from the free outgoing wave; that is, one has a resonance phenomenon in $1+1$ dimensions. Technically, the derivation just given is almost identical to the one of the simple proportionality $\delta g(x,\omega)\propto g(x,-\omega)$ in the preceding paragraph. However, the concept of anomalous points, while central to this paper, can initially be confusing since the definitions of outgoing and incoming waves near (\ref{deffg}) would at first sight seem to be incompatible unless $\omega=0$. One can, if necessary, convince oneself by taking $V(x)=V_0e^{\lambda x}$ ($\lambda>0$) and calculate at least $k$ Born approximations to $f(\omega)$ to see the anomalous point at $\omega'=-ik\lambda/2$ ($k=1$ being the simplest example). Rescaling $f\mapsto{\sf f}$ as in the preceding to remove the anomalous-point divergence, one observes that ${\sf f}(\omega')$ and $f(-\omega')$ agree to any desired order, as they must. Considering the exact solutions available in this\cite{tail} or in the P\"oschl--Teller\cite{susy} case, using standard identities for the Bessel and hypergeometric functions involved one arrives at the same conclusion.

These examples also introduce the following approach: if $V(\lambda^{-1}\ln z)$, continued from $z>0$, is analytic near $z=0$ for certain $\lambda>0$ (the assumption of a decreasing potential tail implying that $V(z{=}0)=0$), the KGE (\ref{KGE}) has a regular singular point\cite{boyce} at $z=0$ in the variable $z=e^{\lambda x}$, and the index equation yields the associated characteristic exponents simply as $\pm i\omega/\lambda$ (from now on $\lambda=1$ not to burden the notation). The Frobenius theory of regular singular points now distinguishes three cases\cite{omega0}. (a) In the generic case $2i\omega\not\in{\bf Z}$, there are two independent generalized power-series solutions, corresponding to the Born series for the outgoing and incoming waves. If $2i\omega\in{\bf Z}$, the small ($\sim z^{i\omega}$ for $\im\omega<0$) solution still is a generalized power series; this is the incoming wave $f(-\omega)$. However, the large solution is of the form $\sum_{j=0}^\infty c_jz^{j-i\omega}+\zeta f(z,-\omega)\ln z$, where $\{c_{j\ge1}\}$ and $\zeta$ have to be calculated case by case. (b) {\em Typically} $\zeta\neq0$, so that there is no generalized power series for the large solution; this is exactly the case of diverging Born series discussed before, and hence corresponds to an anomalous point. Indeed, in the example of the preceding paragraph these were located at $\omega'=-ik/2$. The outgoing wave, defined by analytic continuation, will certainly not be logarithmic in $z$ and hence should correspond to the small solution, as has been deduced already for anomalous points in general. (c) It can also happen that a `miracle'\cite{susy} makes $\zeta=0$ even though $2i\omega\in{\bf Z}$. Hence, in this {\em doubly nongeneric\/} situation, there is a one-parameter family of large generalized power-series solutions, undetermined up to a multiple of the small solution. One of these will correspond to the analytically continued outgoing wave, since in the coordinate $x$ the vanishing of $\zeta$ means that higher-order corrections to $V(x)\sim V_0e^x$ have conspired in a `miraculous' cancellation of the divergence in $f(\omega)$. Thus, case (c) does {\em not\/} correspond to an anomalous point.

While the Frobenius method in itself of course is elementary, the above recapitulation emphasizes its aspects when $\omega$ is not a fixed parameter, but a variable of the theory. Let us complement the discussion by two remarks, one more abstract, the other more computational and concrete.

The first remark is that one can analytically continue the solutions along a circle around $z=0$, and study the so-called monodromy map this generates in the two-dimensional solution space\cite{ince}. For generalized power series the effect follows from the leading coefficient as $z^{\pm i\omega}\mapsto(ze^{2\pi i})^{\pm i\omega}=z^{\pm i\omega}e^{\mp2\pi\omega}$. Hence, in the generic case (a), the monodromy map is a diagonalizable matrix with distinct eigenvalues $e^{\mp2\pi\omega}$. On the other hand, in the anomalous-point case (b) this map is a nontrivial Jordan block (e.g.,\cite{I&P}), because applying it to any large solution yields an image which is not even proportional, as the continuation carries one onto a different branch of the logarithm. The missing eigenvector precisely corresponds to the missing preferred large solution, since both the incoming and outgoing waves are small. This also shows that, if one were to define a preferred `interesting' large solution by other means, such a hypothetical notion of `interesting' cannot be invariant under monodromy, which limits its relevance. Finally, for a `miracle' (c) the monodromy map becomes either $-\openone$ (for half-integer $i\omega$) or $+\openone$ (for integer $i\omega$), so that any solution is an eigenvector. Hence, in the latter case, at the immediate level only the incoming solution is distinguished as the small one, while the preferred increasing solution---the outgoing wave---must be determined by other means.

The second remark concerns precisely the calculation of such miraculous outgoing waves. Generically the coefficients of $f(\omega)=\sum_{j=0}^\infty c_jz^{j-i\omega}$ follow from the recurrence relation $j(j-2i\omega)c_j=\smash{\sum_{p=1}^j}V_pc_{j-p}$, with $V(z)=\sum_{p=1}^\infty V_pz^p$ and $c_0\equiv1$. If $\omega=\omega'=-ik/2$ ($k\in{\bf Z}$), the brute-force approach is to calculate $c_k(\omega)$ from this recurrence for unevaluated $\omega$, and finally investigate the limit $\omega\rightarrow\omega'$. However, even using computer algebra such operations on (big) rational functions are less transparent than those on (large) integers only. Of course, it is standard to verify whether a miracle occurs by checking if $\sum_{p=1}^kV_pc_{k-p}(\omega')$ vanishes. If it does, in this way one has verified {\em that\/} a large outgoing wave exists. However, even to determine {\em which\/} value of $c_k$ corresponds to $f(\omega')$ brute force is not needed. Namely, using de l'Hospital's rule one finds $c_k(\omega')=(-2ik)^{-1}\sum_{p=1}^kV_p\partial_\omega c_{k-p}|_{\omega'}$, where for $1\le p\le k-1$ the {\em numbers} $\partial_\omega c_p|_{\omega'}$ follow from the $\omega$-derivative of the recursion relation, which in the given range of $p$ can never lead to any divergences. Finally, the $c_{j\ge k+1}$ are calculated from the recursion relation without further ado. For the RWE, in the following sections we will obtain complete analytical control, and a double check (which we have performed) using brute-force computer algebra is still feasible for $\ell=2,3$. However, in a more complicated situation in Section~\ref{schw-lim} we shall use an approach analogous to the above to calculate (\ref{c10}).

When it applies (as it does for black holes), Frobenius theory thus yields complete information on anomalous points and miracles. However, other cases are also possible, such as $V(x)\sim xe^{\lambda x}$ (readily handled as $\partial_\lambda[V(x)\sim e^{\lambda x}]$, yielding an index $\nu=2$\cite{susy}), and the responsible exponential tail could even be buried beneath, e.g., an algebraic one; see further Section~\ref{cont-th}.

\subsection{Exact solution of the Regge--Wheeler equation}
\label{solveRWE}

To avoid the transcendental (\ref{x-s}) we write the RWE in terms of $r$,
\bea
  &&[r^2(r{-}1)^2d_r^2+r(r{-}1)d_r\nonumber\\
  &&{}-(r{-}1)\{2(n{+}1)r-3\}+\omega^2r^4]\phi=0\;,
\eeal{RWE}
while (\ref{deffg}) translates to $f(r,\omega)\sim[e(r{-}1)]^{-i\omega}$ and $g(r,\omega)\sim(re^r)^{i\omega}$. At $\omega^2=\Omega^2$, integration of (\ref{defW})--(\ref{SUSYgen}) yields $\xi_1$ as the $\gamma=\gamma_1\equiv0$ case in the exact solution\cite{chandra}
\bea
  \xi(r,\gamma)&=&[(r{-}1)e^r]^{-N\hn/2}\frac{2nr+3}{r}
    \left[1+\gamma I(r)\right],\label{psi-gen}\\
  I(r)&\equiv&\int_1^r\!\frac{dt}{t-1}\,[(t{-}1)e^t]^N\frac{t^3}{(2nt+3)^2}\;,
\eeal{def-I}
readily found by varying the constant in $\xi_1$. The remaining quadrature {\em seems\/} to make $\xi(r,\gamma)$ singular near $2nr+3=0$; however, in fact this cannot happen since the RWE (\ref{RWE}) is regular there, so the singularity must cancel because $n$ and $N$ in (\ref{def-I}) are related as in (\ref{defN}). Motivated by these remarks, one finds that the integral actually is elementary, {\em viz.},
\bea
  I(r)&=&\frac{(2n{-}1)(n{+}1)(N{-}2)!}{3nN^{N+1}}\!\left[\hn e^N
    -e^{Nr}\hn\sum_{j=0}^{N-2}\frac{[N(1{-}r)]^j}{j!}\!\right]\nonumber\\
  &&{}+\frac{e^{Nr}(r{-}1)^{N-1}[2nr^2-(2n{+}3)r+6]}{4n^2N(2nr+3)}\;.
\eeal{res-I}
Various choices of the free parameter $\gamma$ now will correspond to the various physically interesting solutions. We have already defined $\xi_1$ with $\gamma_1=0$ as the SUSY generator (\ref{SUSYgen}). Since incoming waves are readily and uniquely characterized by being decreasing, their determination is immediate. In particular we see that $\xi_1$ itself is incoming from infinity, and from (\ref{psi-gen}) we get in detail $\xi_1=2ng(-\Omega)$. Similarly, the form (\ref{psi-gen}) shows that $\xi_2\equiv\lim_{\gamma\rightarrow\infty}\xi(\gamma)/\gamma=f(-\Omega)/[(2n{+}3)N]$ is incoming from the horizon, i.e., formally $\gamma_2\equiv\infty$. Further, we define $\xi_3$ as the outgoing wave into the horizon $\propto f(\Omega)$, $\xi_4$~as the unique small solution near $r=0$, and $\xi_5$ as the outgoing wave to infinity $\propto g(\Omega)$. It may be helpful to note that for $1\le j\le5$, $\xi_j$ is evaluated in Section~$j$.

For general $\omega$, the RWE (\ref{RWE}) has regular singular points at $r=1$ and $r=0$, and an irregular singular point at $r=\infty$\cite{leaver1}. We shall successively study these three singularities (in fact the behaviors of the solutions at each of them will turn out to be related); upon combination, the results will completely elucidate the situation at the special frequency and in particular prove the claims made in the Introduction.

\section{Near the horizon}
\label{horizon}

At the horizon $r=1$, one has exactly the situation described in Section~\ref{tails} (with $\lambda=1$), so that the index equation has roots $\pm i\omega$. Hence, for $2i\omega=k\in{\bf Z}\backslash\{0\}$, {\em typically\/} the large ($\sim (r{-}1)^{-|\omega|}$) solution is expected to contain a contribution $\sim (r{-}1)^{|\omega|}\ln(r{-}1)$ so that $\omega=-ik/2$ is an anomalous point, as is indeed readily verified for the lowest few $k$. However, one now sees at once that, for any $\ell$, $\omega=\Omega$ is a doubly nongeneric, `miraculous' frequency, for the general solution $\xi(r,\gamma)$ of (\ref{psi-gen}) contains no log-terms at all. Hence, the Born series is finite and convergent, and the outgoing solution $f(\Omega)$ (up to this stage: whatever it may turn out to be) is large so that $\omega=\Omega$ is not an anomalous point. If it seems surprising that, e.g., the simple form of $\xi_1$ alone can lead to such a nontrivial conclusion, one can of course for a given $\ell$ evaluate the first $N-1$ terms of the series for $f(\Omega)$ to see the pertinent cancellation (Section~\ref{tails}) occur.

In contrast to the incoming wave, in the presence of potential tails the outgoing wave is no longer characterized by its leading asymptotic behavior, nor does the solution (\ref{psi-gen}) at a single frequency allow the continuation procedure stipulated below (\ref{deffg}). An elegant indirect method to find the outgoing wave for $r\downarrow1$ is as follows: $\xi$ is outgoing iff $\partial_\omega[(r{-}1)^{i\omega}\xi]$ contains no log-terms. Namely, in the general superposition $\partial_\omega[(r{-}1)^{i\omega}\{Af(\omega)+Bf(-\omega)\}]$ only the second term in square brackets contains $\omega$ in an exponent, so that these log-terms come from the incoming part of $\xi$. Writing things out using the variation-of-constant solution for $\partial_\omega\xi$\cite{jordan}, the criterion becomes that $[r/(r{-}1)]\xi^2(r)$ (the first factor is a Jacobian) should not contain an $(r{-}1)^{-1}$-term. Substituting (\ref{psi-gen}) into this nonlinear function of $\xi$, one finds that this happens iff\cite{incGamma}
\bea
  \gamma=\gamma_3&\equiv&
  -\!\left[\hn\frac{9}{2}\sum_{j=0}^{N-2}\frac{N^{j+1}}{j!}
         +3(2n{+}3)\frac{N^{N+1}}{N!}\right]\!e^{-N}\label{gamma0}\\
  &=&-\frac{28\cdot11093}{3}e^{-8}\qquad\mbox{for $\ell=2$}\;;
\eeal{ell2}
in particular, $\gamma_3$ is a sum of negative terms and thus is nonzero for any $\ell$. Hence, defining $\xi_3\equiv\xi(\gamma_3)=(2n+3)f(\Omega)$, for all $\ell$ one has $\xi_1\not\propto\xi_3\not\propto\xi_2$, where the former inequality means that {\em the RWE has no TTM$_{\rm R}$ at $\Omega$}. Summarizing, we are led to the striking conclusion that in spite of its appealing simple form the SUSY generator $\xi_1$ is not outgoing into the horizon.

The special case (\ref{ell2}) for the `magic number' $\gamma_3$ can be and has been readily verified by using computer algebra to do an eighth-order Born approximation with unevaluated $\omega$, or its refinement described at the end of Section~\ref{tails}. However, for the next-simplest case $\ell=3$ one already needs fortieth order\cite{calg}. Despite this computational demand, these series methods are a reliable way of computing $f(\omega)$, the logarithmic derivative of which can be compared to the one of $g(-\omega)$, the latter being stably obtained by integrating the RWE down from large $r$. In this way we have verified that for $\ell=2$ there is also no TTM$_{\rm R}$ {\em near\/} $\Omega$ on the NIA (say between the nearest two anomalous points, although the rest of the semi-axis is readily checked as well). In fact, for {\em any\/} $\ell$ one readily convinces oneself analytically that the real zero of $f(x,\Omega)$ (cf.\ the asymptotics of $\xi_3$ for $x\rightarrow\pm\infty$) is stable under small imaginary perturbations of $\Omega$. Since for a repulsive potential such a zero means that $f(\omega\approx\Omega)$ cannot simultaneously by incoming from the right, the claim of\cite{and} on the presence of a TTM$_{\rm R}$ is refuted.

The above is enough to classify the RWE$\mapsto$ZE SUSY-transform on the left.  Namely, $\Omega$ has turned out not to be anomalous, while the generator $\xi_1$ is mixed (i.e., a nontrivial superposition of incoming and outgoing). This means that the transformation is of category (b2) in the terminology of Appendix~B3 in\cite{susy}, the result quoted in Section~VI of that paper. We then get the prediction that the `miracle' will not repeat itself for the ZE, where on the left $\Omega$ should be an anomalous point with index $\tilde{\nu}=1$, $\tilde{\xi}_1$~being the only solution with no log-terms\cite{ZElog}; besides, it follows from the general theory that $\tilde{\xi}_1\propto\tilde{{\sf f}}(\Omega)$ is anomalous outgoing. Independent of the precise situation on the right, this means that for the ZE a TTM$_{\rm R}$ at $\Omega$ would simultaneously be an NM, which cannot occur. Hence, {\em also the ZE has no TTM$_{\rm R}$ at $\Omega$}.

Let us conclude this section by pointing out two interesting perspectives on SUSY offered by the preceding. In the first place, in\cite{ZElog} we obtained a very straightforward proof of $\xi_1$ not being outgoing into the horizon---a property referring to the RWE only---by looking at the anomalous point of the ZE. In the second place, in Section~\ref{tails} it has been argued that this very anomalous point is a nuisance numerically near $\Omega$. However, in this section we have seen that the corresponding anomalous point is absent in the RWE, and any numerical result obtained there can subsequently be SUSY-transformed back to the ZE. Thus, SUSY can be used both as an {\em analytical\/} tool for the {\em RWE\/} and as a {\em numerical\/} tool for the {\em ZE}. The theory of\cite{susy} shows that the latter application actually has some generality, namely, as long as the anomalous-outgoing wave is nodeless (as it has to be for repulsive $\tilde{V}$, cf.\cite{alphaOK}).

\section{Near the black-hole singularity}
\label{sing}

The index equation now gives the characteristic exponents $\rho_1=3$, $\rho_2=-1$, so that $\rho_1-\rho_2\in{\bf Z}$. While this is still widely recognized (e.g.,\cite{leaver2}), previous works do not seem to carry out the associated series expansion for a more precise determination of the character of the singularity, i.e., logarithmic/anomalous (case (b) of Section~\ref{tails}) or miraculous (case (c)). The above shows that this determination in fact is decisively simpler than for $r=1$, since the expansion is only needed to fourth order for any $\ell$ and $\omega$. By again merely looking at the {\em form\/} of (\ref{psi-gen}) (cf.\ Section~\ref{horizon}), one sees that at least for $\omega=\Omega$ a miracle happens at $r=0$, since the general solution then contains no $\ln r$-terms. Indeed, actually carrying out the fourth-order expansion readily shows that a power-series solution $\sim r^{-1}$ exists if {\em and only if\/} $\omega^2=\Omega^2$.

Let us now define $\xi_4$ as the unique solution which is finite near $r=0$:
\bea
  \xi_4(r)&=&\left(\frac{e^{-r}}{r{-}1}\right)^{\!\!N\hn/2}\frac{2nr+3}{r}
    \int_0^r\!\frac{dt}{t{-}1}\,\frac{[(t{-}1)e^t]^Nt^3}{(2nt+3)^2}
    \label{psi4}\\
  &=&\xi_1(r)[-I(0)+I(r)]\nonumber\\
  &=&-I(0)\xi_1(r)+\xi_2(r)\;,
\eea
so that $\gamma_4\equiv-I(0)^{-1}$. Obviously $\xi_4\not\propto\xi_1$, since their $r\rightarrow\infty$ asymptotics differ. Also $\xi_4\not\propto\xi_2$ since $I(0)\neq0$, all factors of the integrand having a fixed sign on $(0,1)$.

Could $\xi_4\propto\xi_3=\xi_1+\gamma_3\xi_2$, that is, could the unexpected result $\xi_3\neq\xi_1$ for the outgoing wave into the horizon perhaps be distinguished by its behavior near the black-hole singularity? This is true iff
$I(0)=-\gamma_3^{-1}$. We will now prove that this can{\em not\/} happen; for the required bound, (\ref{res-I}) is less convenient than direct estimation of the integral. Because $N$ is even one has $I(0)>0$, and $I(0)<\frac{1}{9}\int_0^1\!dt\,(1-t)^{N-1}e^{Nt}t^3$; use $\int_0^1\!dt\,(1-t)^ke^{Nt}=(k!/N^{k+1})[e^N-\sum_{j=0}^kN^j\hn/j!]$, doing the terms with $N\le j\le N+2$ exactly. In the remainder use $e^N-\sum_{j=0}^{N-1}N^j\hn/j!>0$, then the contribution of this remainder is seen to be negative, so it can be omitted in the upper bound. After this, the inequality becomes $I(0)<2(N{+}1)/9N^3<1/3N^2$. For $|\gamma_3|$ use $\sum_{j=0}^{N-2}N^j\hn/j!<e^N$, $2n+3<3\sqrt{N}$, and $N!>2N^{N+1/2}e^{-N}$, yielding
\beq
  |\gamma_3|<9N\;.
\eeql{g3ineq}
Combination now gives $I(0)|\gamma_3|<3/N<1$, for $N\ge8$. Hence, $\xi_1$ through $\xi_4$ are all different.

\section{Near \lowercase{$r=\infty$}; the mode situation}
\label{infty}

\subsection{A continuation theorem}
\label{cont-th}

Let us first derive a general result on the KGE (\ref{KGE}): {\em if $V(x)$ is analytic for $x>x_0$ (say), and if its continuation from those $x$ has $V(x)=O(x^{-1-\eta})$ for $|\arg x|<\beta$\cite{symm} and some $\beta,\eta>0$, then $g(x,\omega)\sim1\cdot e^{i\omega x}$ if (A) $|\arg x|<\beta$, (B) $|\pi/2-\arg\omega|<\pi/2+\beta$, and (C) $|\pi/2-\arg\omega x|<3\pi/2$. In particular, $g(\omega)$ has no anomalous points for $|\pi/2-\arg\omega|<\min(\pi/2+\beta,3\pi/2)$}. (The counterpart for $f$ is obvious but not needed here.)

Namely, for $\omega$ obeying (B) the continuation of (\ref{deffg}) can be first carried out in parts of the $x$-plane where it merely amounts to selecting and normalizing the unique small solution, which can never lead to any divergence. This proves the result for $|\pi/2-\arg\omega x|<\pi/2$, which in a second step can be extended at fixed $\omega$ by trivially solving (\ref{KGE}) in a region where $V$ is negligible; condition (C) (implied by (A) and (B) if $\beta\le\pi/2$) arises because the asymptotic form for $g$ cannot be continued in this way from a region where it dominates across an anti-Stokes line.

In applications, the second step would typically take the form of continuation back to the physical region $x>x_0$. The rationale for the two-step procedure is seen to be that continuations in $x$ are easier than those in $\omega$, since for the former one has the linear ordinary differential equation (\ref{KGE}) at one's disposal. For an example, calculating a few terms in the Born series shows the theorem in action for $V(x{>}x_0)=e^{-\lambda x}\cos(\mu x)$\cite{susy}, which has $\beta=\arctan(\lambda/|\mu|)$.

\subsection{Application to the Regge--Wheeler equation}
\label{appRWE}

Turning to the RWE, the above procedure is best carried out in terms of $r$. Since $V(r)$ vanishes if $|r|\rightarrow\infty$, one has $\beta=\infty$ so that $g$ has no anomalous points, its asymptotics being as below (\ref{RWE}) for $|\pi/2-\arg\omega r|<3\pi/2$\cite{noproblem,negx}. This should now be compared with the asymptotics of $\xi$, for which (\ref{psi-gen}) and (\ref{res-I}) show that
\bea
  \xi(r,\gamma)&\sim&
  2n\left[1+\gamma\frac{(2n{-}1)(n{+}1)(N{-}2)!e^N}{3nN^{N+1}}\right]
    (re^r)^{\!-N\hn/2}
  \nonumber\\ &&{}+\frac{\gamma}{2nN}(re^r)^{\!N\hn/2}
\eeal{psi-asy}
in the entire $r$-plane. In particular, for both $g_{\rm l}$ (which has $\arg\Omega=3\pi/2$) and $g_{\rm r}$ (with $\arg\Omega=-\pi/2$) one can make the comparison with (\ref{psi-asy}) in the region $\re r\le0$, where the asymptotics are not dominated by the second term on the r.h.s., as well as for the original $x>0$. This of course reproduces that $\xi(\gamma)$ is incoming from infinity for $\gamma=0$, but also shows that it is outgoing {\em for both branches of $g(\omega)$} iff
\beq
  \gamma=\gamma_5\equiv\frac{3nN^{N+1}e^{-N}}{(1-2n)(n+1)(N-2)!}\;,
\eeql{gamma5}
in which case $\xi_5\equiv\xi(\gamma_5)=(\gamma_5/2nN)g_{\rm l/r}(\Omega)$, in accordance with the general statement (see\cite{susy}, Appendix~B2) that if $g_{\rm l}$ and $g_{\rm r}$ coincide they must be real\cite{confrac}. Hence, the discontinuity index $\mu\ge1$ at $\Omega$.

This statement on $\mu$ (dealing with analytic continuations in $\omega$ and with $r\rightarrow\infty$) has thus been related to the `miracles' near $r=1$ and $r=0$ (dealing with continuations for small $r$ at a fixed $\omega$), since the latter make that $\xi(r,\gamma)$ has the same asymptotics for $\arg r<-\pi/2$ and $\arg r>\pi/2$. Note that for a qualitatively similar but otherwise arbitrary potential, the miracle is a `leftist' and the vanishing of $\alpha(\omega)$ a `rightist' concept, so that the two would be completely unrelated. Of course, the RW potential is anything but arbitrary\cite{chandra}, being of the form $V(x)=W^2(x)-d_xW(x)+\Omega^2$, as follows at once from (\ref{x-s}), (\ref{VRW}), and (\ref{defW}) or indeed from SUSY in general.

Obviously $\xi_1\not\propto\xi_5\not\propto\xi_2$, where the former inequality means that $\Omega$ indeed is not anomalous for the RWE on the right, and the latter that {\em the RWE has no TTM$_{\rm L}$ at $\Omega$}. Two questions remain, namely, whether possibly $\gamma_5=\gamma_3$ or $\gamma_5=\gamma_4$; the answer is negative both times. For the latter (Section~\ref{sing} showed that the outgoing wave into the horizon was not distinguished near $r=0$, but could perhaps the outgoing wave to infinity be?), this is immediate since (\ref{res-I}) for $I(0)=-\gamma_4^{-1}$ shows that $\gamma_4^{-1}-\gamma_5^{-1}$ is a sum of positive terms, hence nonzero. For the former, using methods as above (\ref{g3ineq}) (in particular, $N!<\sqrt{8}N^{N+1/2}e^{-N}$) leads to $|\gamma_5|>(\sqrt{3}/4)N^2$. If $\ell\ge3\Rightarrow N\ge40$, this implies $|\gamma_5|>9N>|\gamma_3|$ (cf.\ (\ref{g3ineq})), while for the remaining case $\ell=2$ the inequality $|\gamma_5|=e^{-8}2^{24}\hn/135>|\gamma_3|$ (cf.\ (\ref{ell2})) is immediate. Hence, $\xi_5\not\propto\xi_3$ and {\em the RWE has no QNM at $\Omega$}, even though the positive discontinuity index $\mu\ge1$ in itself would have allowed this. Thus, the five solutions we have identified---while of course not independent---are all different, and the surprise is rather that there are not six of them, i.e., that $g_{\rm l}(\Omega)=g_{\rm r}(\Omega)$, given that the outgoing wave for the Schwarzschild black hole definitely has a branch cut in general (see\cite{leaver2,tail,noBorn} and Section~\ref{indices}).

\subsection{SUSY and the Zerilli equation}
\label{SUSYZE}

The mere absence of anomalous points in $g$ for the RWE (cf.\ above (\ref{psi-asy})) already implies that the SUSY generated by $\xi_1$ is of type (a1) on the right in the terminology of\cite{susy}. The general theory then shows that $\tilde{\xi}_1\propto\tilde{g}_{\rm l}(\Omega)=\tilde{g}_{\rm r}(\Omega)$. Since we have shown in Section~\ref{horizon} that it is also anomalous outgoing on the left, {\em $\tilde{\xi}_1$ thus is a QNM=TTM$_{\rm L}$ of the ZE\/} (of course, this once more shows that the ZE has no TTM$_{\rm R}$ at $\Omega$).

Because $\mu\ge1$ already in the original system, the equality of $\tilde{g}_{\rm l/r}$ in the above can immediately be sharpened to $\tilde{\mu}\ge2$ by the elementary SUSY relationship
\beq
  \tilde{\alpha}(\omega)=\frac{\Omega-\omega}{\Omega+\omega}\hp\alpha(\omega)\;.
\eeql{SUSYalpha}
However, to determine the orders of the modes one needs the full apparatus of\cite{susy}. Its Eqs.~(B6), (B7) show that the QNM Wronskian $\tilde{J}_{\rm q}(\omega)$ is regular near $\Omega$, while the TTM Wronskian $\tilde{J}_{\rm t}(\omega)$ has a first-order zero there and a simple pole at $-\Omega$. Accounting for the anomalous point of $\tilde{f}$, this means that {\em both the QNM and the TTM$_{\rm L}$ of the ZE furnished by $\tilde{\xi}_1$ are simple\/}---intuitively plausible, since the RWE has no QNMs/TTM$_{\rm L}$s to double the ones generated by SUSY. The proportionalities involved are elucidated as $\tilde{\xi}_1=\tilde{\xi}_2=\tilde{\xi}_3/\gamma_3=\tilde{\xi}_4=\tilde{\xi}_5/\gamma_5=\tilde{f}(-\Omega)/(2n{+}3)= -i(2n{+}3)\gamma_3^{-1}\tilde{{\sf f}}(\Omega)=\tilde{g}(\Omega)/(2n)$.

Incidentally, since $\tilde{V}(|r|{\rightarrow}\infty)$ vanishes just as well as the RW potential, $\tilde{g}$ has the same asymptotics as established above (\ref{psi-asy}) for $g$. Analogously continuing $\tilde{\xi}_1(r)$ to $\re r\le0$ then reproduces that this function is outgoing on the right, without having to calculate the general solution of the ZE at $\Omega^2$\cite{ZElog} and independent of SUSY. This also shows why the ZE having $\tilde{\alpha}(\Omega)=0$ in the absence of a `miracle' does not contradict the discussion below (\ref{gamma5}), relating the two. Namely, in this case $\tilde{g}$, being a TTM$_{\rm L}$, is small and hence single-valued near the branch point of the general solution.

For readers mainly interested in results, it should be pointed out that this subsection {\em inter alia\/} proves that the ZE retarded Green's function has a pole at $\omega=\Omega$, the statement of which is completely independent of the intricacies of anomalous outgoing waves.

\subsection{Discontinuity indices}
\label{indices}

The above has established the mode situation at the special frequency. It will now be shown that the derived inequalities for the discontinuity indices are in fact realized as equalities, i.e., that one has $\mu=1$, which implies $\tilde{\mu}=2$ by (\ref{SUSYalpha}). To this end, define $g_1(\omega)\equiv\partial_\omega g(\omega)$\cite{jordan} and make the variation-of-constant Ansatz $g_1(\Omega)=g(\Omega)h(\Omega)$\cite{pole-h}. One finds
\beq
  d_xh(x,\Omega)=-2\Omega\hp g^{-2}(r,\Omega)
  \int_{r_0}^r\!dt\,\frac{t}{t-1}\hp g^2(t,\Omega)\;,
\eeql{dxh}
and to get the correct asymptotics $d_xh(x,\Omega)\rightarrow i$ also if $r$ is continued to the left half-plane, one should set $r_0=-\infty$. However, for $g_{1\rm r}(r,\Omega)$ ($g_{1\rm l}(r,\Omega)$) the asymptotics are imposed if $r$ is continued into the second (third) quadrant, so that for $r>0$ one must put $t\mapsto t+i\epsilon$ ($t\mapsto t-i\epsilon$) in the integrand of (\ref{dxh}). Evaluating the residues at $t=1$ and $t=0$, one finds in detail
\beq
  \delta\!\int_{-\infty}^{r>1}\!\!\!\!dt\,\frac{t\hp\xi_5^2(t)}{t-1}=
  2\pi i\!\left[\frac{2}{N}(\gamma_3{-}\gamma_5)
    +9\!\left(\frac{\gamma_5}{\gamma_4}-1\right)^{\!\!2}\right]
\eeql{Delta}
(see Section~\ref{tails} for the definition of $\delta$). In particular, since it has already been shown in Section~\ref{appRWE} that $|\gamma_5|>|\gamma_3|$, the first term in brackets of (\ref{Delta}) (which is due to the pole at $t=1$, and hence anticipated to be $\propto\gamma_3-\gamma_5$) is positive as well as the second so that $\delta g_1(\Omega)\neq0$ and $\mu=1$\cite{mZE}. Eqs.~(\ref{dxh}) and (\ref{Delta}) will have an interesting application in Section~\ref{discuss}.

\section{Leaver series}
\label{sec:leaver}

In\cite{confrac} we already touched upon the relation of this work to the first Leaver series
\beq
  f(r,\omega)\propto\left(\frac{r^2e^r}{r-1}\right)^{\!\!i\omega}
  \sum_{j=0}^\infty c_j\left(\frac{r-1}{r}\right)^{\!\!j}.
\eeql{leaverf}
However, at least as interesting a connection exists with the second Leaver series,
\bea
  g(r,\omega)&\propto&r^3\left(\frac{e^r}{r-1}\right)^{\!\!i\omega}
    \sum_{j=0}^\infty c_j(1-2i\omega)_j\nonumber\\
    &&{}\times U(3{+}j{-}2i\omega,5,-2i\omega r)
\eeal{leaverg}
in terms of the same $c_j$ as in (\ref{leaverf}), and one of the more `irregular' cases (integer second argument) of the irregular confluent hypergeometric function\cite{erdelyi},
\bea
  U(&&\kappa,5,z)=
    -\frac{1}{24\Gamma(\kappa-4)}\biggl[M(\kappa,5,z)\ln z\nonumber\\
  &&{}+\sum_{p=0}^\infty\frac{(\kappa)_p}{(5)_p}\frac{z^p}{p!}
    \{\psi(\kappa{+}p)-\psi(1{+}p)-\psi(5{+}p)\}\biggr]\nonumber\\
  &&{}+\frac{6}{\Gamma(\kappa)z^4}M_4(\kappa{-}4,-3,z)\;,
\eeal{hyperU}
where $M_4$ is the series to four terms of the regular confluent hypergeometric function $M$, and $\psi=\Gamma'\!/\Gamma$.

To see the connection, let $2i\omega=L=1,2,\ldots$, so that the Pochhammer symbol reduces to $(1-L)_j=(-)^j(L{-}1)!/(L{-}j{-}1)!$ for $0\le j\le L-1$, and zero for $j\ge L$. Typically this zero is compensated by the anomalous-point divergence in $c_j$, so that the series still involves $U(5,5,z)=e^z\Gamma(-4,z)$ and higher, which have logarithmic branch points at $z=0$ (cf.\ (\ref{hyperU})). However, if $\omega=\omega_{\rm m}$ is a `miraculous' frequency, the sum in (\ref{leaverg}) truncates:
\bea
  g(r,\omega_{\rm m})&\propto&r^3\left(\frac{e^r}{r-1}\right)^{\!\!L/2}
  \sum_{j=0}^{L-1}c_j\frac{(-)^j}{(L{-}j{-}1)!}\nonumber\\
  &&{}\times U(3{+}j{-}L,5,-Lr)\;,
\eeal{leaverm}
in which only those $c_j$'s enter which are the same for all solutions which are large near $r=1$, so that in particular for $L=N$ one has $c_j=(1/j!)\*{d_u^j|}_{u=0}\exp\{N/(u{-}1)\}[2n(1{-}u)^N+3(1{-}u)^{N+1}]$ (combine (\ref{psi-gen}) and (\ref{leaverf})). The functions $U$ involved in (\ref{leaverm}) are evaluated from (\ref{hyperU}) as\cite{U34}
\bea
  U(2,5,z)&=&z^{-2}+4z^{-3}+6z^{-4}\;,\nonumber\\
  U(1,5,z)&=&z^{-1}+3z^{-2}+6z^{-3}+6z^{-4}\;,\nonumber\\
  U(-k,5,z)&=&\sum_{p=0}^k\left(\begin{array}{c}k\\ p\end{array}\right)
    (p+5)_{k-p}(-)^{k+p}z^p\nonumber\\
  &&(k=0,1,\ldots)\;,
\eea
Since $U(2,5,z)$ and $U(1,5,z)$ are linearly independent and since $c_{L-1}$ and $c_{L-2}$ cannot both vanish for a nonzero sequence $\{c_j\}$ because of the recursion
\bea
  &&(j+1)(j+1-L)c_{j+1}\nonumber\\ &&{}-[2j^2+(2{-}4L)j+2L(L{-}1)+2(n{+}1)-3]c_j
  \nonumber\\ &&{}+(j^2-2Lj+L^2-4)c_{j-1}=0\;,
\eeal{recur}
one concludes that $g(r,\omega_{\rm m})$ is a large single-valued solution near $r=0$. In Section~\ref{sing} it has already been established that this can only happen for $\omega_{\rm m}^2=\Omega^2$, so that {\em$\Omega$ is the only `miraculous' frequency of the RWE}. It should be emphasized that this is a nontrivial statement on the singularity near $r=1$, and that the above proof involved an analysis both near $r=0$ and $r=\infty$.

Thus, $L=N$ is the only case which actually occurs, and hence the r.h.s.\ of (\ref{leaverm}) should evaluate to $2(n{+}1)N^{-3}\xi_5(r)$ as given by (\ref{psi-gen}), (\ref{res-I}) and (\ref{gamma5}), where the constant of proportionality (for the $c_j$ below (\ref{leaverm})) has been determined from the leading asymptotics for $r\rightarrow\infty$. Verification amounts to comparing two polynomials of degree $N+1$ with rational coefficients, which has been done explicitly for $\ell=2$. For general $\ell$, the stated outcome of the sum in (\ref{leaverm}) can be proved by downward induction in the power of $r$, using the recursion (\ref{recur}) upward; however, we omit the laborious details.

The expression (\ref{leaverg}), (\ref{hyperU}) is also of interest for nonmiraculous $\omega=-i|\omega|$. Namely, it shows that in the Leaver series for $\delta g$ one has
\beq
  \delta U(3{+}j{-}2i\omega,5,-2i\omega r)=
  \frac{i\pi M(3{+}j{-}2i\omega,5,-2i\omega r)}{12\Gamma(j{-}2i\omega{-}1)}\;.
\eeql{dUM}
The relevance of this latter relation is not so much that the {\em regular\/} $M$ is easier to handle than the $U$ function (which it is), but rather that it points to a computational scheme for $\delta g$ directly and not as an exponentially small difference between the $g_{\rm l/r}$. In particular, the proportionality derived in Section~\ref{tails} means that one can evaluate $\alpha(\omega)$ at any convenient $r>1$, and a method of calculation which avoids substracting nearly equal large numbers would enable one to look for, e.g., possible simplifications as $r\rightarrow\infty$. However, the properties of the (non-truncating) sum (\ref{leaverg}) itself on the NIA are still under investigation\cite{mak}. In its turn, this last remark also motivates the explicit verification of (\ref{leaverm}) in the previous paragraph.

\section{Kerr holes}
\label{kerr}

\subsection{Kerr-hole special modes}
\label{kerr-spec}

Both the exact solution for the stationary metric and the separated wave equations for its linearized perturbations can be generalized from the spherically symmetric Schwarzschild hole to the axisymmetric, rotating Kerr hole. Presently we investigate the issues addressed in the preceding sections in this wider context, and start by establishing some notation. Expanding the pertinent effective scalar field as
\beq
  \psi(\omega,r,\theta,\phi)=\sum_{\ell=2}^\infty\sum_{m=-\ell}^\ell e^{im\phi}
  S(\theta,\omega)\hp R(r,\omega)
\eeq
in Boyer--Lindquist coordinates, the radial equation reads\cite{grav}
\bea
  \Delta^2d_r\!\hn\left(\frac{d_rP}{\Delta}\right)+\left[
  \frac{K^2+is(1{-}2r)K}{\Delta}+4is\omega r-\lbar\right]\!P&=&\nonumber\\
  \left[\Delta\!\left(d_r{-}i\frac{sK}{2\Delta}{+}\frac{1{-}2r}{\Delta}\right)
  \!\!\left(d_r{+}i\frac{sK}{2\Delta}\right)\!+3is\omega r-\lbar\right]\!P
  &=&0\;,\nonumber
\eea

\vspace{-7mm}
\beq
\eeql{rad-TE}
where we introduced the auxiliary quantities
\bea
  \Delta&=&r^2-r+a^2\nonumber\\
    &=&(r-r_+)(r-r_-)\;,\\
  r_\pm&=&\frac{1\pm b}{2}\;,\\
  b&=&\sqrt{1-4a^2}\;,\\
  K&=&(r^2+a^2)\omega-am\;,
\eea
and where $P=\Delta^{1+s/2}R$. In (\ref{rad-TE}), $0\le a<\half$ is the black-hole rotation, and $s=\pm2$ is the spin of the field. The radial field $P$ depends on $(r,\omega,\ell,m,a)$ and on the sign of $s$, but this will only be indicated where necessary. The angular separation constant $\lbar$ is an eigenvalue of the equation for $S(\theta,\omega)$, cf.\ (\ref{lbar-exp}) below, and different eigenvalues correspond to different angular-momentum sectors. Since $\lbar$ obeys the symmetries $\lbar_m^*(\omega)=\lbar_{-m}^\phs(-\omega^*)$, $\lbar_m(\omega)=\lbar_{-m}(-\omega)$, and $\lbar_s=\lbar_{-s}$ (besides, $\lbar(\omega,a)=\lbar(\omega a)$), from (\ref{rad-TE}) it follows that $P_m^*(\omega)=P_{-m}^\phs(-\omega^*)$ and $P_{ms}(\omega)=P_{-m,-s}(-\omega)$; an advantage of using $P$ instead of $R$ is that this makes the latter relation manifest. Moreover, there is the considerably deeper symmetry between $P_s$ and $P_{-s}$ furnished by the Teukolsky--Starobinsky identities (which follow by combining the finite-$a$ generalization of (\ref{RWZ-BP}) below with its inverse for opposite $s$, see also\cite{grav}). Since the special modes (\ref{Pspec}) below are in the kernel of the corresponding symmetry operator (cf.\ (\ref{Apm})) we do not give the formulas; however, the {\em existence\/} of this symmetry will be used at the end of Section~\ref{specQNM}.

Where the mere existence of the {\em ordinary\/} differential equation (\ref{rad-TE}) is already remarkable, this holds {\em a fortiori\/} for its exact solvability if $\omega=\Omega(a)$ is a zero of the so-called Starobinsky constant\cite{notreal}
\bea
  {\cal S}&=&\lbar^2(\lbar{+}2)^2
  +8\lbar[5\lbar(am\omega{-}a^2\omega^2)+6(am\omega{+}a^2\omega^2)]
  \nonumber\\ &&{}+36\omega^2+144(am\omega{-}a^2\omega^2)^2\;.
\eeal{star}
Namely, in that case the special modes $P=\cal P$ are given by\cite{chandra}
\bea
  {\cal P}&=&\biggl[is\Omega r^3+\lbar r^2-i\frac{sQ}{8\Omega}r-2a^2
    +{\ts\frac{2}{3}}\lbar\left(a^2{-}\frac{am}{\Omega}\right)\nonumber\\
    &&\quad{}-\frac{\lbar{+}2}{24\Omega^2}Q\biggr]\!\left(\frac{r-r_+}{r-r_-}
    \right)^{\!\!i\frac{sam}{2b}}e^{-is\Omega x/2}\,,\label{Pspec}\\
  Q&=&\lbar(\lbar{+}2)+12(am\Omega{-}a^2\Omega^2)+3is\Omega\;,
\eea
with the tortoise coordinate
\beq
  x=r+\frac{r_+}{b}\ln(r{-}r_+)-\frac{r_-}{b}\ln(r{-}r_-)
\eeql{kerr-x}
mapping $r_+<r<\infty$ to $x\in{\bf R}$. In view of the aforementioned symmetries, one can take $\im\Omega<0$ for consistency with the preceding sections.

When assessing the physical significance of the solutions (\ref{Pspec}), in comparison with the RWE/ZE case one encounters two technical subtleties, which are readily dealt with. Writing (\ref{rad-TE}) as a KGE in the $x$-coordinate for $\sqrt{r^2+a^2}P/\Delta$, one finds that the ensuing potential is (i) complex and frequency dependent, and (ii) tending to a nonzero constant if $x\rightarrow-\infty$, and having an $O(x^{-1})$ tail for $x\rightarrow\infty$. (i) If $V=V(x,\omega)$, besides the outgoing waves (\ref{deffg}) one also has to define incoming waves $F,G$ (temporary notation), being solutions of (\ref{KGE}) at $\omega$ with $G(x{\rightarrow}\infty,\omega)\sim e^{-i\omega x}$ for $\im\omega<0$ and continued analytically from there, and analogously for $F$. In the frequency-independent case these become simply $G(\omega)=g(-\omega)$ etc., but in general $G$ and $g$ are unrelated. Fortunately, for (\ref{rad-TE}) one has $G_{ms}(\omega)=g_{-m,-s}(-\omega)$. (ii) A potential having a nonzero spatial limit can be handled by a shift in frequency; in particular, for (\ref{rad-TE}) it is customary to define $\tilde{\omega}=\omega-ma/r_+$. The $O(x^{-1})$ potential tails give rise to a power-law prefactor in the wave-function asymptotics, not unlike the one occurring for the RWE when $g$ is considered as a function of $r$, cf.\ below (\ref{RWE}); these power laws will cause no problems here.

The character of the special modes on the right is now easily determined. With our convention on $\Omega$, it follows from (\ref{Pspec}) that ${\cal P}_2\equiv{\cal P}_{s=2}$ is decaying and hence incoming from $x=+\infty$, while ${\cal P}_{-2}$ is outgoing by the continuation argument of Section~\ref{cont-th}. Near the horizon $r=r_+$, the characteristic exponents of (\ref{rad-TE}) read $\rho_{1(2)}=1\pm(1+isr_+\tilde{\omega}/2b)$. For the special modes one thus has $\rho_1-\rho_2=2+isr_+\tilde{\Omega}/b$, with $\Omega$ a zero of (\ref{star}). For $a\!\downarrow\!0$ this becomes $\rho_1-\rho_2\rightarrow2+sN\!/2\in{\bf Z}$, but one does not expect the difference to keep this value for nonzero $a$. Indeed, the expansion (\ref{Omega-exp}) below shows that $\rho_1-\rho_2$ is non-integer for all sufficiently small nonzero $a$. While we have not attempted a proof, it seems very unlikely that, in some $(\ell,m)$-sector, the complex function $\rho_1(a)-\rho_2(a)$ would assume a real integer value exactly on $0<a<\half$. Thus, for nonzero rotation, the algebraically special frequency $\Omega(a)$ is actually generic in the terminology of Section~\ref{tails}. Since (\ref{Pspec}), (\ref{kerr-x}) show that the modes $\cal P$ are generalized power series corresponding to $\rho_2$ (equivalently, they are eigenfunctions of the monodromy map near $r_+$ with eigenvalue $e^{2\pi i\rho_2}$), this means that ${\cal P}_2$ is outgoing into the horizon {\em and hence a TTM$_{\rm R}$\/} for $a>0$. Similarly, ${\cal P}_{-2}$ is incoming from the horizon {\em and thus a TTM$_{\rm L}$}; this follows already from its being decreasing on the left, so that for ${\cal P}_{-2}$ the condition $a>0$ is not needed.

For $a=0$, in which case the above analysis is inconclusive for ${\cal P}_2$, the radial equation (\ref{rad-TE}) does not reduce to the RWE or ZE, but to the Bardeen--Press equation instead; the explicit mapping between them reads\cite{grav}
\bea
  P&=&r^3\biggl[\biggl\{\left(W-\frac{N}{2}\right)
  (2nr+2\pm1)-\frac{1}{r}-is\omega\biggr\}\nonumber\\ &&\hphantom{r^3\biggl[}
  \times\left\{d_x-\frac{is\omega}{2}\right\}
  +W^2\mp d_xW-\frac{N^2}{4}\biggr]\phi\label{RWZ-BP}\\
  &\equiv&A_\pm\phi\;,
\eea
where the upper (lower) signs should be chosen if $\phi(r,\omega)$ is a solution of the RWE (ZE). Calculating the Born series for $f$ for (\ref{rad-TE}) with $\ell=2$ (so that $\lbar=4$, cf.\ (\ref{lbar2}) below), at $\Omega=-4i$ one finds a divergence (like for the ZE, but here in sixth order) if $s=-2$ but not if $s=+2$ (like for the RWE, but here the relevant cancellation occurs in tenth order), so that $\Omega$ is anomalous (miraculous) for $s=-2$ ($+2$). However, ${\cal P}_2$ is a nontrivial superposition of outgoing and incoming waves---again involving the `magic number' 11093, cf.\ (\ref{ell2})---in marked contrast to the situation for any nonzero $a$. Thus, the unexpected findings of Section~\ref{horizon} are not an artifact of the mapping (\ref{RWZ-BP})---a kind of $\omega$-dependent SUSY transform---but intrinsic to the Schwarzschild limit, which apparently has to be highly singular. To investigate this further, one should study (\ref{rad-TE})--(\ref{Pspec}) for small $a$; this is done in the next subsection.

\subsection{Schwarzschild limit}
\label{schw-lim}

It should be explained why the $s=2$ special mode behaves differently for $a=0$ than for all other $a$. This is not a consequence of a singularity in (\ref{Pspec}) itself, which in the limit tends smoothly to
\begin{mathletters}
\bea
  {\cal P}_2^{(0)}(r)&=&\frac{1}{6n}[8n^2(n{+}1)r^3+12n^2r^2+18nr+9]\nonumber\\
    &&{}\times e^{-Nx/2}\;,
  \label{P2a0}\\
  {\cal P}_{-2}^{(0)}(r)&=&-\frac{2n}{3}[2(n{+}1)r^3-3r^2]e^{Nx/2}\;,
\eeal{P-2a0}\end{mathletters}%
where one should note the similarity with (\ref{VZ}) and (\ref{VRW}) respectively, and where the quantitative relation to the special solutions of the RWE/ZE reads
\begin{mathletters}
\bea
  A_+\xi_1&=&2nN{\cal P}_2^{(0)}\delta_{s2}\;,\label{Ap}\\
  A_-\tilde{\xi}_1&=&-\frac{N}{2n}{\cal P}_{-2}^{(0)}\delta_{s,-2}\;.
\eeal{Am}\label{Apm}\end{mathletters}%

Since it has been shown in Section~\ref{kerr-spec} that ${\cal P}_2$ is outgoing on the left and that $\Omega(a)$ is generic for nonzero $a$, this implies
\bea
  \lim_{a\downarrow0}\lim_{\omega\rightarrow\Omega(a)}f_2(\omega,a)&=&
  \frac{2n{\cal P}_2^{(0)}}{(2n{+}3)(N{+}1)}\nonumber\\
  &=&\frac{A_+\xi_1}{(2n{+}3)N(N{+}1)}\;,
\eeal{limit1}
where $f_2\equiv f_{s=2}$ has been normalized as
\beq
  f_2(x{\rightarrow}{-}\infty,\omega,a)\sim1\cdot e^{-i\tilde{\omega}x}\;,
\eeq
and where the constant of proportionality has been obtained by combining this last relation with (\ref{P2a0}) and (\ref{Ap}). On the other hand, by the very definition of an outgoing wave, taking the limits in the opposite order yields an outgoing result, i.e.,
\bea
  \lim_{\omega\rightarrow\Omega}\lim_{a\downarrow0}f_2(\omega,a)&=&
  f_2(\Omega,a{=}0)\nonumber\\
  &=&\frac{A_+\xi_3}{(2n{+}3)N(N{+}1)}\;,
\eeal{limit2}
where the constant of proportionality is the same as in (\ref{limit1}) because $\xi_1$ and $\xi_3$ have equal increasing parts, cf.\ (\ref{psi-gen}).

To see how these noncommuting limits arise, set $f_2(r,\omega)=e^{-i\tilde{\omega}r_+}b^{i\tilde{\omega}r_-/b}\sum_{j=0}^\infty c_j(r-r_+)^{j-i\tilde{\omega}r_+/b}$ (i.e., $c_0=1$), upon which the $c_j$ follow form the recursion
\beq
  \alpha_jc_j+\beta_jc_{j-1}+\gamma_jc_{j-2}+\delta_jc_{j-3}+\epsilon_jc_{j-4}
  =0\;,
\eeql{c-recur}
\bea
  \alpha_j&=&b^2j\Bigl(j-2-2i\tilde{\omega}\frac{r_+}{b}\Bigr)\;,\nonumber\\
  \beta_j&=&b\Bigl(j-1-i\tilde{\omega}\frac{r_+}{b}\Bigr)
    \Bigl(2j-7-2i\tilde{\omega}\frac{r_+}{b}\Bigr)\nonumber\\
    &&{}+4r_+\tilde{\omega}(r_+\omega{-}i)+b(4ir_+\omega{-}\lbar)\;,\nonumber\\
  \gamma_j&=&\Bigl(j-2-i\tilde{\omega}\frac{r_+}{b}\Bigr)
    \Bigl(j-5-i\tilde{\omega}\frac{r_+}{b}\Bigr)\nonumber\\
    &&{}+2r_+\omega(2r_+\omega{+}\tilde{\omega})+6ib\omega-\lbar\;,\nonumber\\
  \delta_j&=&4\omega(r_+\omega{+}i)\;,\nonumber\\
  \epsilon_j&=&\omega^2\;.
\eea
It is now readily seen {\em that\/} one has noncommuting limits, because
\bea
  \alpha_{N+2}&=&(N{+}2)[2ima-Na^2-2(i\omega{-}N\!/2)]+O(ma^2)
  \nonumber\\ &&{}+O(a^3)+O[(\omega{+}iN\!/2)a]+O[(\omega{+}iN\!/2)^2]
  \nonumber\\
\eeal{alphaN2}
implies that $c_{N+2}$ (and hence higher $c_j$) are sensitive to the ratio of $\omega+iN\!/2$ and $a$. To calculate {\em which\/} are the limiting functions, one has to iterate (\ref{c-recur}) up to $c_{N+1}$, up to the error terms indicated in (\ref{alphaN2}). This in fact is a calculation of the type outlined at the end of Section~\ref{tails}, except that here one has {\em three\/} numerical coefficients in the recursion relation---the lowest-order one plus the leading $\omega+iN\!/2$ (linear) and $a$ (linear for $m\neq0$, quadratic for $m=0$) corrections. In terms of\cite{seidel}
\beq
  h(\ell)=\frac{(\ell^2-m^2)(\ell^2-4)^2}{2(\ell-\half)\ell^3(\ell+\half)}\;,
\eeql{hell}
the required expansion of the separation constant reads
\bea
  \lbar&=&\ell(\ell+1)-2-\left(2+\frac{8}{\ell(\ell+1)}\right)\!ma\omega
  \nonumber\\ &&{}+[h(\ell{+}1){-}h(\ell)]a^2\omega^2+O[(a\omega)^3]
  \label{lbar-exp}\\
  &=&4-{\ts\frac{10}{3}}am\omega+\left({\ts\frac{10}{21}{-}\frac{10}{189}}m^2
     \right)\!a^2\omega^2\nonumber\\
     &&{}+O[(a\omega)^3]\qquad\mbox{for $\ell=2$}\label{lbar2}\;,
\eea
and for $\ell=2$ some computer algebra yields
\beq
  c_{10}=\frac{3633559(4{-}i\omega)+37411744ima+42040064a^2}
  {1786050(i\omega-4-ima+4a^2)}\;,
\eeql{c10}
where both the numerator and the denominator are accurate up to the error terms indicated in (\ref{alphaN2}). From (\ref{c10}) it is immediate that
\beq
  \lim_{\omega\rightarrow\Omega}\lim_{a\downarrow0}c_{10}
  =-\frac{3633559}{1786050}\equiv c_{10}^{(3)}\;.
\eeql{lim2c}
To also find the limit corresponding to (\ref{limit1}), substitute (\ref{lbar-exp}) into (\ref{star}), yielding its roots as
\bea
  \Omega(a)&=&\Omega_0+\Omega_1ma+\Omega_{20}a^2+\Omega_{21}m^2a^2\nonumber\\
    &&{}+O(ma^3)+O(a^4)\label{Omega-exp}\\
  &=&-4i+{\ts\frac{32}{3}}ma+{\ts\frac{128}{189}}i(65m^2{-}18)a^2\nonumber\\
    &&{}+O(ma^3)+O(a^4)\qquad\mbox{for $\ell=2$}\;,\label{Om-l2}\\
  \Omega_0&=&-\half iN\;,\nonumber\\
  \Omega_1&=&{\ts\frac{2}{3}}nN\;,\nonumber\\
  \Omega_{20}&=&{\ts\frac{1}{3}}inN^2
    \frac{(\ell{-}3)(\ell{+}4)}{(2\ell{-}1)(2\ell{+}3)}\;,\nonumber\\
  \Omega_{21}&=&i\frac{4n^3[21\ell^2(\ell{+}1)^2+2\ell(\ell{+}1)+12]}
    {27(2\ell{-}1)(2\ell{+}3)}\;,\nonumber
\eea
which in its own right is perhaps a new result for the asymptotics of the special frequencies, agreeing well with the numerical values in\cite{onozawa} for $\ell=2$ and the smallest $a$. Note that $\Omega_{20}$ vanishes for $\ell=3$, and has opposite signs for $\ell=2$ and $\ell\ge4$. Using (\ref{Om-l2}) in (\ref{c10}) now yields
\beq
  \lim_{a\downarrow0}\lim_{\omega\rightarrow\Omega(a)}c_{10}
  =-\frac{69632}{893025}\equiv c_{10}^{(1)}\;,
\eeql{lim1c}
where in particular one obtains the same limit both for $m\neq0$ and $m=0$ even though these two cases are quite different for (\ref{c10}) and (\ref{Om-l2}) individually. Since the normalized function $\lim f_2(\omega,a)$ is a solution of the Bardeen--Press equation, which is completely specified by its coefficient $c_{N+2}$, and since for $\ell=2$ the values (\ref{lim2c}), (\ref{lim1c}) are readily checked to correspond to (\ref{limit2}) and (\ref{limit1}) respectively, the above clarifies the mechanism for the singularity of the Schwarzschild limit, being the zero in $\alpha_{N+2}$.

\subsection{Special QNMs}
\label{specQNM}

We have seen in the preceding subsection that $f_2(\omega,a)$ is wildly varying near $(\omega{=}{-}iN\!/2,a{=}0)$. Thus, if near this point $f_2$ can tend both to $A_+\xi_1$ and $A_+\xi_3$ (up to normalization), can it also tend to $A_+\xi_5$? After all, the latter is merely a linear superposition of the former two. Thus, one should look for solutions of
\beq
  c_{10}(\omega,a)=c_{10}^{(5)}\equiv
  \Bigl(1{-}\frac{\gamma_5}{\gamma_3}\Bigr)c_{10}^{(1)}
  +\frac{\gamma_5}{\gamma_3}c_{10}^{(3)}=-\frac{2166784}{893025}
\eeq
in the region of validity of (\ref{c10}). Indeed one finds that $\lim_{a\downarrow0}\lim_{\omega\rightarrow\omega_{\rm q}(a)}c_{10}(\omega,a)=c_{10}^{(5)}$ for
\bea
  \omega_{\rm q}&=&-4i-\frac{33078176}{700009}ma+\frac{3492608}{41177}ia^2
  \nonumber\\ &&{}+O(ma^2)+O(a^4)\;.
\eeal{om-q}
This result has a significant interpretation: to leading order, for $\omega=\omega_{\rm q}(a)$ the outgoing function into the horizon $f_2$ is also outgoing to infinity, so that (\ref{om-q}) {\em represents a family of QNMs of (\ref{rad-TE}) for $(\ell{=}2,a{\neq}0)$, branching from the special frequency $\Omega$ in the Schwarzschild limit}.

Clearly, one is interested in the generalization of (\ref{om-q}) to arbitrary $\ell$. This can be performed in closed form, since (\ref{c10}), obtained in the above using computer algebra, can be reconstructed from its known limits together with the expansion (\ref{alphaN2}). In fact, considerable cancellation occurs in the answer: while for reference we give the generalizations of (\ref{lim1c}) and (\ref{lim2c}) as
\beq
  c_{N+2}^{(1)}=-\frac{4n^2(8n^5{+}12n^4{+}18n^3{-}13n^2{-}18n{-}45)}
  {27(2n{+}3)(N{+}1)(N{+}2)!}\left(\frac{N}{2}\right)^{\!\!N}
\eeq
[from (\ref{P2a0}) and (\ref{limit1})] and
\beq
  c_{N+2}^{(3)}=
  c_{N+2}^{(1)}+\frac{6\gamma_3e^N}{(2n{+}3)^2N(N{+}1)^2(N{+}2)}
\eeq
[from (\ref{limit2}) with $A_+\xi_3=A_+\xi_1+\gamma_3A_+\xi_2$ and $A_+\xi_2=6e^{N\!/2}(r{-}1)^{N\!/2+2}/[(2n{+}3)(N{+}1)(N{+}2)]+O[(r{-}1)^{N\!/2+3}]$ by the explicit form (\ref{RWZ-BP}) for $s=2$ and the three leading coefficients of $\xi_2$ as yielded by (\ref{psi-gen}), (\ref{def-I})] respectively, also without these explicit values one can derive the comparatively simple final result
\bea
  \omega_{\rm q}&=&-\half iN
  +\frac{\gamma_5-\Omega_1\gamma_3}{\gamma_5-\gamma_3}ma
  +\frac{N\gamma_5+2i\Omega_{20}\gamma_3}{2(\gamma_5-\gamma_3)}ia^2
  \nonumber\\ &&{}+O(ma^2)+O(a^4)\;.
\eeal{om-qg}
In particular, while the $O(a)$-coefficient thus keeps its negative sign for $\ell>2$, the quadratic coefficient, governing the behavior of the special QNMs with $m=0$ and positive in (\ref{om-q}), is seen to become negative if $\ell\ge4$.

For a powerful and instructive check on the preceding, we shall presently rederive the $m\neq0$ case (i.e., only the first two terms on the r.h.s.) of (\ref{om-qg}) from the radial equation (\ref{rad-TE}) for $s=-2$. Namely, by the Teukolsky--Starobinsky identities this latter equation is physically equivalent to (in particular, has the same QNM spectrum as) the $s=2$ equation as long as $\omega\neq\Omega(a)$, a condition which (\ref{om-qg}) fulfills for $a\neq0$. To do so, we investigate in which direction in the $(\omega,a)$-space the QNM (\ref{P-2a0}) can be perturbed so that it continues to satisfy outgoing boundary conditions. In its turn, this calculation buttresses our interpretation of ${\cal P}_{-2}^{(0)}$ (and $\tilde{\xi}_1$ for the ZE) as {\em bona fide\/} QNMs even though they are decreasing on the left; however, this last property will be technically important in the calculation below.

Thus, let $P_{\rm q,-2}(a)$ be the branch of special QNMs, with $\omega=\omega_{\rm q}(a)=-iN/2+\omega_{\rm q1}ma+O(a^2)$ and $P_{\rm q,-2}(0)={\cal P}_{-2}^{(0)}$, and set $d_aP_{\rm q,-2}(a)|_{a=0}\equiv im{\cal P}_{-2}^{(0)}h_{\rm q}$. Differentiation of (\ref{rad-TE}) yields
\bea
  d_rh_{\rm q}(r)&=&\frac{r(r{-}1)}{{{\cal P}_{-2}^{(0)}}^2(r)}\int^r\!\!dt\,
  \frac{{{\cal P}_{-2}^{(0)}}^2(t)}{t^3(t{-}1)^3}\hp B(t)\;,\label{drhq}\\
  B(t)&=&N\omega_{\rm q1}t^4+4\omega_{\rm q1}t^3
  +\biggl(\frac{8n}{3}{-}6\omega_{\rm q1}\biggr)t^2\nonumber\\
  &&{}+4\biggl(1{-}n{-}\frac{n^2}{3}\biggr)t-2\;.
\eeal{Bs}
Our limitation to linear perturbations in $a$ enables one to set $r_+,b=1+O(a^2)$ etc.\ in the derivation of (\ref{drhq}) and (\ref{Bs}), without which these expressions would have been considerably more complicated. For $P_{\rm q,-2}$ to be outgoing to infinity, on has to set $\int^r\!dt=\int_{-\infty}^r\!dt$ in (\ref{drhq}), by a repetition of the arguments in Section~\ref{infty}. Since the integrand is entire, one does not need to specify a path; physically this means that our lowest-order calculation is free from branch-cut complications. On the other hand, for outgoing waves into the horizon, one sees from the Born series that $d_rh(r)$ should have a residue $1-\omega_{\rm q1}$ at $r=1$. Writing $\int^r\!dt=K_{\rm q}+\int_1^r\!dt$ and expanding the integrand near $t=1$, the residue criterion is seen to imply
\beq
  K_{\rm q}=\frac{8n^2}{9c}(1-\omega_{\rm q1})\;,
\eeql{Kq}
with $c$ being the residue at $r=1$ of
\bea
  &&\frac{e^{-Nr}}{(r{-}1)^{N-1}r^3[2(n{+}1)r-3]^2}=
  -\frac{2ne^{-Nr}[(N{-}n)r{+}n{+}1]}{9(r{-}1)^{N+2}r}\nonumber\\ &&{}-d_r
  \frac{e^{-Nr}[2n(4n{+}1)r^2+3(1{-}2n)r-3]}{18(r{-}1)^{N+1}r^2[2(n{+}1)r-3]}\;,
\eea
that is, $c=-\gamma_3/27$. For the perturbation to be outgoing to both sides, i.e., to represent a QNM, we now have to solve $\omega_{\rm q1}$ from
\beq
  K_{\rm q}=\int_{-\infty}^1\!\!\!dt\,
  \frac{{{\cal P}_{-2}^{(0)}}^2(t)}{t^3(t{-}1)^3}\hp B(t)\;,
\eeq
where a tedious but elementary calculation yields the r.h.s.\ as $8n^2(3\omega_{\rm q1}{-}2nN)/\gamma_5$. Combination with (\ref{Kq}) now gives $\omega_{\rm q1}=(\gamma_5{-}\frac{2}{3}nN\gamma_3)/(\gamma_5{-}\gamma_3)$, i.e., consistent with (\ref{om-qg}), which is what we set out to show.

\section{Discussion}
\label{discuss}

While over the years the singular nature of the extreme Kerr limit $a\uparrow\half$ for rotating black holes has been increasingly well understood, the present work shows that also the Schwarzschild limit $a\downarrow0$ is highly delicate. Besides thus pointing out the problems, considerable progress has also been made towards their solution, culminating in (\ref{om-qg}) for the special QNMs, involving a perhaps beautiful combination of $\gamma_3$ (Section~\ref{horizon}), $\gamma_5$ (Section~\ref{infty}), and $\Omega_1,\Omega_{20}$ (Section~\ref{schw-lim}). However, besides this technical control, it would be desirable to also have a physical understanding. In particular, the total-transmission property of the `algebraically special' solutions is related in\cite{chandra} to the vanishing of the Weyl scalars $\Psi_0$ or $\Psi_4$ as these occur in the Newman--Penrose formalism of general relativity, an aspect outside this article's scope. At present it is not yet clear why the argument fails for the Schwarzschild hole.

Thus, a Newman--Penrose analysis would be a welcome complement to our wave-equation study on the conceptual side. On the practical side---the calculation of black-hole spectra---the prospects for numerical progress on or near the NIA\cite{mak} seem good, even though the present work refutes essentially all numerical claims about the nature of the RWE at $\Omega$ made previously. {\em First}, we have proved that the QNM $\Omega'$ reported in\cite{leaver1} cannot possibly coincide with $\Omega$. A spurious $\Omega'$ could for instance be found if the numerical code {\em ad hoc\/} compensates for an anomalous-point divergence in the $c_j$ of (\ref{leaverf}) which at $\Omega$ actually is absent; however, comparison of the results of Section~\ref{specQNM} with the data in\cite{onozawa} in fact suggests the existence of $\Omega'$, see below. If an imaginary $\Omega'$ would be distinct from, but numerically very close to $\Omega$ (even though it is doubtful whether the data of\cite{leaver1} itself support this conclusion, cf.\ the Introduction), this would imply that $\alpha(\omega)$ has two very nearby zeros, which can be investigated numerically.

{\em Second}, the analysis of\cite{and} has some serious problems. While the paper does point out (below Table~1) that the pattern of (anti-)Stokes lines near the horizon central to its analysis actually collapses on the imaginary $\omega$-axis---where its data are reported---the matter is not pursued to conclude that on that axis analytic continuations in $r$ are not suitable to find outgoing waves to start with\cite{negx}. Rather, on that axis the generic/anomalous/miraculous distinction is crucial, but the latter cannot be treated by the WKB ideas used in\cite{and}, which are more appropriate for studying $|\omega|\rightarrow\infty$ than for numerics at a finite~$\omega$. Indeed, Refs.\cite{LYT,dirt} find that (for $\re\omega\neq0$) continuations in $r$ can be used as a numerical tool independently of any WKB-type approximations. Remarkably, in each angular-momentum sector $(\ell,m)$, of the two one-parameter families---labeled by $(a,s{=}{\pm}2)$---of TTMs claimed in\cite{chandra}, the only one which has been comfirmed numerically (the TTM$_{\rm R}$ at $a=0$) thus is the one which in fact does not exist!

{\em Third}, it has been found in Section~\ref{specQNM} that in each $\ell$-sector the Schwarzschild special frequency $\Omega(0)$ splits into {\em one\/} $m$-multiplet of special QNMs as the rotation is increased from zero. In particular, only for $m<0$ do we predict these QNMs to branch into the fourth $\omega$-quadrant, in contrast with the claim in\cite{onozawa}. However, it should be noted that the coefficients in (\ref{om-q}) are quite large, so that the Schwarzschild limit is approached only for extremely small $a$; as yet no numerical data exist for this regime. In fact, it has been shown in Section~\ref{kerr} that for the $s=-2$ radial equation, in this limit the special QNM and TTM$_{\rm L}$ become identical (like for the ZE), while for $s=2$ the single (in each $(\ell,m)$-sector) QNM and TTM$_{\rm R}$ cancel each other (like for the RWE). Thus, while\cite{onozawa} correctly cautions for subtleties of the Schwarzschild limit, its precise nature cannot be elucidated from numerics alone. The preceding does not account for the ninth branches of $(\ell{=}2,m{=}1,2)$ Kerr QNMs which\cite{onozawa} reports for moderately small $a$, and which have slightly smaller $\abs{\im\omega}$ than their apparent $m=-1,-2$ counterparts, the latter being accounted for by (\ref{om-q}). Since in the Schwarzschild limit all Kerr QNMs have to merge as $m$-multiplets by spherical symmetry, this strongly suggests that the former QNMs merge with their $m<0$ mirror images into an imaginary $\Omega'$, with $\abs{\Omega'}\lesssim\abs{\Omega(0)}$\cite{other-poss}. The $m=0$ branches emerging from $\Omega(0)$ and $\Omega'$ then supposedly move towards each other along the NIA, until they collide and subsequently move away from this axis. This scenario is consistent with the sign of the $O(a^2)$-term in (\ref{om-q}); the remark below (\ref{om-qg}) then hints that $\abs{\Omega'}>\abs{\Omega(0)}$ for $\ell\ge4$. Additional numerical work is needed to explore these fascinating possibilities; also, for a general understanding of the Schwarzschild limit it is unfortunate that the figures in\cite{onozawa} do not show some Kerr modes branching from the high-damping series of Schwarzschild QNMs.

A detailed computation of $\alpha(\omega)$ as proposed in relation to\cite{leaver1} would also be of more general interest. Namely, in Section~\ref{sec:leaver} we have proved that $\Omega$ is the only `miraculous' frequency of the RWE, and the question suggests itself whether it likewise is the only zero of the discontinuity function $\alpha$. It turns out that this cannot be the case in general. For small $\omega$, one can use the Born analysis of\cite{tail}. True to form, the Schwarzschild potentials are an exceptional case (a major conclusion of\cite{tail}) in that the leading-order contribution to $\alpha(\omega)$ expected generically vanishes. Fortunately the actual leading-order behavior still follows from the first Born approximation, as
\beq
  \alpha(\omega)=(-)^{\ell+1}2\pi\omega+\mbox{h.o.t.}\;,
\eeql{alpha1}
incidentally for the ZE as well as for the RWE as required by (\ref{SUSYalpha}). Eq.~(\ref{alpha1}) can now be compared to the upshot of Section~\ref{indices}, {\em viz.},
\bea
  \frac{d[i\alpha(\Omega)]}{d(i\omega)}&=&
  8\pi n^2\!\left[2N\frac{\gamma_3-\gamma_5}{\gamma_5^2}
    +9N^2\!\left(\frac{1}{\gamma_4}-\frac{1}{\gamma_5}\right)^{\!\!2}\right]\!
    \label{alpha2}\\
  &>&0\;.\nonumber
\eea
In both calculations one finds that the $x$-dependence in $\alpha(\omega)\equiv\delta g(x,\omega)/g(x,-\omega)$ cancels, as it must. Comparison of (\ref{alpha1}) and (\ref{alpha2}) now shows that {\em the parity of zeros (counting their multiplicities) of $\alpha(-iz)$ on $(0,N/2)$ is $(-)^\ell$, so that in particular for {\em odd} $\ell$ the function $\alpha(-iz)$ must have at least one zero on the stated interval}. Numerical work should urgently verify this last conclusion, as well as (\ref{alpha1}), (\ref{alpha2}) in general, also in the context of the claim in\cite{leaver1} that $\re i\Omega'<N/2$ for $\ell=2$, but $>N/2$ for $\ell=3$. Further, it should be investigated whether WKB methods or those outlined at the end of Section~\ref{sec:leaver} can establish the sign of $i\alpha(\omega)$ analytically also for $i\omega\rightarrow\infty$. In any case, in view of (\ref{SUSYalpha}), in the latter limit those signs should be opposite for the RWE and ZE, unlike the situation for $i\omega\downarrow0$.

The way in which throughout this article analytic continuations in $r$ and in $\omega$ have turned out to be related points to some interesting function theory in ${\bf C}^2$. While these pure-mathematical aspects have not been pursued, as their consequence and almost coincidentally we have arrived at a fairly complete theory of anomalous points and miracles in general, that is, independent of the specifics of the potentials (\ref{VRW}), (\ref{VZ}). Namely, in practice most potential tails will be either exponential, covered by the Frobenius method in Section~\ref{tails}, or algebraic, covered by the two-step theorem of Section~\ref{cont-th}. In particular, the latter, while comparatively intractable in other contexts because of their diverging Born series, have turned out not to lead to anomalous points.

A similar degree of generality has not yet been achieved for describing merging and canceling modes, as occurring in Section~\ref{kerr}, or in\cite{susy} for the P\"oschl--Teller model. A model-independent study of merging modes is available as the Jordan-block perturbation theory of\cite{jordan}, but there one starts by {\em assuming\/} the presence of a higher-order mode, which is subsequently split. In particular, it is not known whether the cancellation phenomenon is restricted to long-range potentials. The study of such issues will be facilitated by generalizing to long-range potentials the concepts of a QNM norm and more generally a generalized inner product between two-component vectors (composed of both fields and their associated momenta), which are known to be highly effective for finite-range potentials\cite{RMP}.

The `practical' significance of this work mainly lies in its results on the `special' family of QNMs at small $a$, and in the guidance it gives to any subsequent numerical investigations. Namely, there seems to be no reason why the rotation of an astrophysical black hole should ever vanish exactly. Furthermore, even for real-time numerical experiments one does not predict the `special' ZE QNM to be {\em independently\/} observable: since the QNMs are not complete\cite{comp}, the highly damped non-oscillating signal of the latter is in principle indistinguishable from the late-time tail caused by the branch cut in $g$\cite{tail}, consistent with the ZE's SUSY-equivalence to the RWE, which has been shown in Section~\ref{appRWE} to have no `special' QNM.

However, apart from the dependence of our small-$a$ results on the analysis {\em at\/} $a=0$, there is no further justification required to study the RWE and ZE closely, given that they have the same status in gravity as the Schr\"odinger and Dirac equations for the hydrogen atom have in quantum mechanics. Indeed, given its global singularity structure, especially the RWE is the logical next step up in complexity from the hypergeometric equation encountered in the hydrogen problem\cite{leaver2}. In fact, a look at a figure of the Schwarzschild spectrum (e.g.,\cite{leaver1}) shows that the special frequency is completely central, as the place from which the low- and high-damping branches somehow emerge. Hence, one has the hope that an improved understanding of $\Omega$ will lead to further insight into the spectrum in general, for instance by relating the two numerically observed branches to the two potential tails\cite{KY}, or more ambitiously, by providing accurate and general asymptotic formulae.

\section*{Acknowledgment}

I thank P.T. Leung, Y.T. Liu, K.W. Mak, W.M. Suen, H. Verlinde, C.W. Wong and in particular K. Young for discussions and comments on the preprint, and the ITP Beijing as well as the Universities of Amsterdam and Delft for their hospitality. The work is supported in part by the Hong Kong Research Grants Council (grant CUHK 4006/98P).

\appendix

\section{An afterthought on $\alpha$}
\label{alpha-app}

Our aquaintance with $\alpha$ has proceeded stepwise. Defined originally as a set of proportionality constants in (\ref{def-alpha}), immediately below at least locally on the NIA the {\em function\/} $\alpha(\omega)$ was considered, focusing on the orders of its zeros. Subsequently, in Section~\ref{discuss} the function was studied globally on this axis, and simple ideas of real analysis were used to make a statement about the zeros of $\alpha$. The logical elaboration is to continue $\alpha$ analytically from the NIA, observing that (\ref{def-alpha}) then yields the continuation of $g_{\rm l}(\omega)$ ($g_{\rm r}(\omega)$) to the fourth (third) quadrant and beyond. One attractive property of $\alpha$ is that it is invariant under finitely supported perturbations of $V$. In fact this last result can be sharpened at once: a perturbation $\sim e^{-\lambda x}$ cannot change $\alpha(\omega)$ for $|\im\omega|<\lambda/2$, and subsequently this invariance can be continued to the whole $\omega$-plane.

Let us presently derive a further important property: {\em if the potential $V(x)$ satisfies the decay property of Section~\ref{cont-th} with $\beta=\infty$ and if in addition it is single-valued near $x=+\infty$ (so that $V(x{>}x_0)=\sum_{j=2}^\infty c_jx^{-j}$), then $\alpha(\omega)$ is entire}. Thus, for these potentials the definition (\ref{def-alpha}) is even more fortunate than was apparent at first sight: from a two-variable multiple-valued function it creates a one-variable non-branching function.

For a proof, supplement $g(x,\omega)$ with a branch cut on the negative real $x$-axis (in addition to the cut in $\omega$ on the NIA), set $\vec{g}(x,\omega)\equiv(g(x,\omega),g(x,-\omega))^{\rm T}$, and define the monodromy map by
\beq
  \vec{g}(-\Lambda{-}i\epsilon,\omega)=
  {\cal M}(\omega)\hp\vec{g}(-\Lambda{+}i\epsilon,\omega)
\eeql{def-M}
($\epsilon,\Lambda>0$). Elementary properties are ${\cal M}_{jk}(\omega)={\cal M}_{3-j,3-k}(-\omega)$ ($j,k=1,2$) and ${\cal M}^*(\omega)={\cal M}^{-1}(-\omega^*)$; this last relation involves a matrix inversion because for complex $x$ one has $g^*(x,\omega)=g(x^*,-\omega^*)$, so that under complex conjugation the roles of $-\Lambda\pm i\epsilon$ are reversed. Furthermore, the two-step theorem of Section~\ref{cont-th} implies that ${\cal M}_{22}(\omega)=1$ in the fourth quadrant. To relate $\alpha$ to $\cal M$, observe that for a given $\omega$, the two-step theorem yields the asymptotic form of $g$ for a range of $\arg x$ obeying condition (C) of Section~\ref{cont-th}. Subsequently, $g$ can be continued at fixed $\omega$ to the physical region $x>0$ by use of (\ref{def-M})---effectively adding a third step to the two-step theorem---and comparison with (\ref{def-alpha}) allows one to read off $\alpha$. Thus, for instance for $\re\omega<0$ one finds
\beq
  \alpha(\omega)={\cal M}_{12}(\omega)\;,
\eeql{alpha-M}
making the relation between $\omega$- ($\alpha$) and $x$-continuations ($\cal M$) maximally explicit\cite{cuts}. Now let $\omega=i|\omega|$, and consider the equality $g(-\Lambda{-}i\epsilon,\omega{-}\eta)={\cal M}_{11}(\omega{-}\eta)\hp\*g(-\Lambda{+}i\epsilon,\omega{-}\eta)+{\cal M}_{12}(\omega{-}\eta)\hp\*g(-\Lambda{+}i\epsilon,\eta{-}\omega)$ for $\eta\!\downarrow\!0$. By the preceding one has ${\cal M}_{11}(\omega{-}\eta)=1$, while in the limit $g(-\Lambda{-}i\epsilon,\omega)=g^*(-\Lambda{+}i\epsilon,\omega)$ and $g(-\Lambda{+}i\epsilon,\eta{-}\omega)\rightarrow g_{\rm r}(-\Lambda{+}i\epsilon,-\omega)$. Since the latter function is $\sim e^{i\omega\Lambda}$ and thus real for $\epsilon\!\downarrow\!0$, it follows that $\re\lim_{\eta\downarrow0}{\cal M}_{12}(\omega{-}\eta)=0$. Combination with (\ref{alpha-M}) and the symmetry $\alpha(-\omega^*)=-\alpha^*(\omega)$ then shows that $\lim_{\eta\downarrow0}[\alpha(\omega{-}\eta)-\alpha(\omega{+}\eta)]=0$, so that $\alpha$ has no branch point at the origin. Since a Born calculation (cf.\ (\ref{alpha1})) shows that $\alpha$ is also bounded near the origin, and since under the stated conditions on $V$ the function cannot have singularities for finite $\omega$ (cf.\ Section~\ref{cont-th}), the derivation is complete.

The above is a result of some power: while for algebraic potential tails the large-$x$ expansion of $g$ furnished by the Born series diverges, one now sees that under the stated conditions on $V$ the same series used as a small-$\omega$ expansion of $\alpha$ necessarily {\em con\/}verges. Also, in this situation one has explicit expressions for {\em all\/} branches of $g$; for instance, $g(\omega e^{-2\pi i})=[1+\alpha(\omega)\hp\alpha(-\omega)]g(\omega)+\alpha(\omega)\hp g(-\omega)$ if $\re\omega>0$. Unfortunately, the only available exactly solvable example $V(x{>}x_0)=\sigma(\sigma{+}1)x^{-2}$ has $\alpha(\omega)=2i\sin(\pi\sigma)$, i.e., an $\omega$-independent constant which is not particularly instructive. Note that (pending a closer investigation of the convergence of (\ref{leaverg})) the result (\ref{dUM}) suggests that in the RW case the cut $\delta g$ itself is entire, so that $\alpha$ would have the same branching properties as~$g$; there is no contradiction, for (\ref{VRW}) is not single-valued at infinity in the $x$-variable.

For potentials which cause a cut in only one of the outgoing waves (such as (\ref{VRW}) and (\ref{VZ})), one has the result that zero-modes can only be located at a zero of $\alpha$ on the NIA (Ref.\cite{susy} and cf.\ Section~\ref{SUSYZE}). The above development prompts the question whether this result can be generalized to other frequencies. However, the most obvious of such generalizations cannot be true: under finitely supported perturbations of $V$, off-axis QNMs will vary continuously\cite{dirt} while the zeros of $\alpha$ will not move at all, so that the two cannot coincide in general. In fact, such QNM perturbations are interesting already on the NIA: since the zero of $\alpha$ which coincides with the original QNM will be invariant, it follows that the QNM cannot be shifted along the NIA. Hence, it either has to move away from the axis into a physical branch of the Green's function, and hence split by symmetry, or move into an unphysical branch and hence effectively disappear. Clearly, in the presence of branch cuts in the outgoing wave(s), the Green's function on the NIA and beyond into its unphysical branches merits considerable further study, both for its fundamental interest and for the potential of QNM poles on these unphysical branches to cause numerical artifacts.

\end{multicols} 
 

\begin{references}
\bibitem{out-in}For consistency with the general treatment of (\ref{KGE}), we use an observer-centred definition of `outgoing into the horizon'; however, several other authors use a hole-centred definition `outgoing from the horizon,' in which case on the left the notions of outgoing and incoming are reversed. The last remaining ambiguity, arising since (\ref{KGE}) involves $\omega^2$ only and since `outgoing' at $\omega$ obviously is equivalent to `incoming' at $-\omega$ (see further Section~\ref{tails}), is resolved in the standard manner by choosing $\im\omega<0$ (modes with real $\omega$ will not occur).
\bibitem{NM}At least the possibility that $\xi_1$ could itself be a QNM can be rejected even without considering its concrete form (\ref{defW})--(\ref{SUSYgen}): $\xi_1$ being an `ordinary' QNM, i.e., one not coinciding with other types of modes and hence increasing at both ends, translates into $\tilde{\xi}_1$ being an NM. However, below (\ref{VZ}) this has been ruled out already.
\bibitem{openq}In fact, whether the phenomenon of higher-order modes is restricted to the NIA is still an open question.
\bibitem{alphaOK}Since $g(x,-\omega)$ cannot have a real zero for a purely repulsive potential, in this case---which includes (\ref{VRW}) and (\ref{VZ})---the definition $\alpha(\omega)\equiv\delta g(x,\omega)/g(x,-\omega)$ is not only correct for generic $x$, but in fact for all real $x$. Compare with an analogous situation in\cite{jordan}.
\bibitem{omega0}A fourth case, the odd man out $\omega=0$, is known to be exceptional\cite{susy} but does not concern us here.
\bibitem{incGamma}One can express the sum for $\gamma_3$ in terms of the incomplete gamma function $\Gamma(N-1,N)$; however, doing so has no advantage here.
\bibitem{calg}Computer algebra should also be useful to calculate the Leaver series\cite{leaver1} for $f$ (an expansion in $(r{-}1)/r$), which converges more rapidly than the Born series, and for all $1<r<\infty$. Namely, for rational $r$ and rational-complex $\omega$ one can then use exact integer arithmetic at least for the first few terms, which largely cancel\cite{LYT}.
\bibitem{ZElog}This can be verified without performing laborious series expansions: a second solution of the ZE at $\Omega^2$, independent of $\tilde{\xi}_1$, is provided by $(2n/N)\tilde{g}(r,-\Omega)=\xi_1^{-1}(r)\int_r^\infty\!dt\,[t/(t{-}1)]\hp\xi_1^2(t)$, which indeed has a logarithmic branch point at $r=1$, the integrand having the same nonzero residue there which caused $\gamma_3\neq0$ in (\ref{gamma0}). With the general solution of the ZE at $\Omega^2$ thus at one's disposal, it is straightforward to study, e.g., its additional singularity at $2nr+3=0$. However, this will not be taken up here.
\bibitem{negx}In contrast, $V(x{\rightarrow}{-}\infty)$ only has $\beta=\pi/2$, so that we could not have determined $f(\Omega)$ by continuation nor, of course, ruled out anomalous points for $f$ on the imaginary axis.
\bibitem{confrac}Since Leaver's continued-fraction criterion for $g$\cite{leaver1} amounts to demanding that $\xi(r,\gamma)(re^r)^{-N\hn/2}$ be analytic near $r=\infty$ (cf.\ (\ref{leaverf})), comparison with (\ref{psi-asy}) at once reproduces $\gamma=\gamma_5$. The criterion is thus seen to be a shorthand for the asymptotic matching of the main text.
\bibitem{noBorn}Unfortunately, a direct comparison with the methods of\cite{tail} is not possible. Namely, the error made in the first Born approximation is roughly $\sim V(r)^2$. However, even for large $r$ these algebraically small corrections in general will not be small for the cut $\delta g(r,\omega)\propto g(r,-\omega)$ (cf.\ Section~\ref{tails}; this exact proportionality is not stressed in\cite{tail}), which is {\em exponentially\/} small in $r$ unless $\omega\rightarrow0$ simultaneously with or before $r\rightarrow\infty$. It is precisely this latter case which is studied in\cite{tail}, while we need results at finite $\omega=\Omega$ (however, see Section~\ref{discuss}). Thus, the statement in\cite{tail} below (4.13) should be taken to mean that the leading small-$\omega$ coefficient of $\delta g(\omega)$ is found exactly, not the whole function.
\bibitem{pole-h}Since the asymptotics for $\xi_5(x)\propto g(x,\Omega)$ have opposite sign for $x\rightarrow\pm\infty$, $g(x,\Omega)$ has a single real zero, so that the Ansatz leads to a real pole in $h(x,\Omega)$. However, this will cause no problems.
\bibitem{mZE}The analogous exercise for the ZE of course leads to $\int^r\!dt\,[t/(t{-}1)]\hp\tilde{g}^2(t,\Omega)$, and since $\tilde{g}(\Omega)\propto\xi_1^{-1}$ this is precisely (\ref{def-I}). From (\ref{res-I}) it now follows that $(2nr{+}3)I(r)$ is entire so that indeed $\delta\tilde{g}_1(\Omega)=0$, consistent with $\tilde{\mu}=2$.
\bibitem{U34}Curiously, thus the truncated series does not involve $U(3,5,z)$ and $U(4,5,z)$ although these are single-valued as well. The changes in character of $U$ occur at the zeros of the third coefficient in $U(k{-}1,5,z)-(2k{-}5{+}z)U(k,5,z)+k(k{-}4)U(k{+}1,5,z)=0$, a relation which can also be used when doing the sum in (\ref{leaverm}) by induction, as outlined in the main text.
\bibitem{notreal}We use $\cal S$ in lieu of the more conventional $|{\cal C}|^2$, since $\cal S$ in general is not real unless $\omega$ is. Likewise, Ref.\cite{chandra} uses `complex conjugated' for some object pairs which are only so for real~$\omega$.
\bibitem{other-poss}Other possibilities would be that the `extra' branches of Kerr-QNMs terminate at a finite small $a$, or disappear towards $\omega=-i\infty$; as yet there is no analytical support for the existence of an $\Omega'$.
\bibitem{cuts}From (\ref{alpha-M}) one sees that nontrivial monodromy (i.e., a nondiagonal matrix $\cal M$) is accompanied by a cut in $\omega$ (i.e., $\alpha\neq0$). There is no contradiction with Section~\ref{tails}, where for the anomalous case (b) one has nontrivial monodromy in the absence of an $\omega$-cut in $g$, since the spatial variables involved ($z=x^{-1}$ vs. $z=e^x$) are different.
\end{references}
\end{document}